\documentclass[10pt,a4paper]{article}
\usepackage[utf8]{inputenc}
\usepackage{booktabs,caption}
\usepackage{threeparttable}
\usepackage{amsmath}
\usepackage{amsfonts}
\usepackage{amssymb}
\usepackage{graphicx}
\usepackage[capposition=top]{floatrow}
\usepackage{tabto}
\usepackage[ruled]{algorithm2e}
\usepackage[left=3.25cm,right=3.25cm,top=3cm,bottom=3cm]{geometry}
\usepackage{setspace}
\usepackage{pdflscape}
\usepackage{adjustbox}
\usepackage[title]{appendix}
\usepackage{hyperref}
\hypersetup{pdftex,colorlinks=true,allcolors=blue}

\begin{document}

\begin{titlepage}
	\begin{center}
		\vspace*{1cm}
			
		\Large
		\textbf{Central Bank Communication and the Yield Curve: A Semi-Automatic Approach using Non-Negative Matrix Factorization}
		
		\vspace{1.25cm}
		
		\textbf{Ancil Crayton}\footnote{I am a PhD candidate in the School of Economics at University College Dublin. I can be contacted by email at ancil.crayton@ucdconnect.ie. I would like to thank my supervisor, Karl Whelan, for his immense help during the development of this paper. I would also like to thank Derek Greene, Oana Peia, and Clemens Struck for their comments and advice which also helped shape this paper. Additionally, I would like to thank Stephen Hansen for his advice and help during my visiting stay at Oxford. All mistakes present are my own.}\\
		\textit{University College Dublin} \\
		\vspace{1.5cm}
		
		\today
		
		\vspace{3.0cm}

		\textbf{Abstract}\\
	\end{center}
Communication is now a standard tool in the central bank's monetary policy toolkit. Theoretically, communication provides the central bank an opportunity to guide public expectations, and it has been shown empirically that central bank communication can lead to financial market fluctuations. However, there has been little research into which dimensions or topics of information are most important in causing these fluctuations. We develop a semi-automatic methodology that summarizes the FOMC statements into its main themes, automatically selects the best model based on coherency, and assesses whether there is a significant impact of these themes on the shape of the U.S Treasury yield curve using topic modeling methods from the machine learning literature. Our findings suggest that the FOMC statements can be decomposed into three topics: (i) information related to the economic conditions and the mandates, (ii) information related to monetary policy tools and intermediate targets, and (iii) information related to financial markets and the financial crisis. We find that statements are most influential during the financial crisis and the effects are mostly present in the curvature of the yield curve through information related to the financial theme.
\medskip


\medskip


\end{titlepage}

\section{Introduction}
\onehalfspacing
Communication is becoming a prominent tool in the central bank's monetary policy toolkit. In the trend that central banks are becoming more transparent, we find central banks communicating with the public more frequently and through various avenues. These sources of information reveal not only policy decisions, but also important views about the state of the economy that are taken into consideration when making monetary policy decisions and signaling future policy decisions. The views reflected and communicated in these releases are important for setting the public's expectations about the future state of the economy.

Important to our study is the lack of research systematically identifying the themes within communication sources released by central banks. Assuming efficiency, markets are not expected to adjust simply to an arbitrary release of a statement, but adjustment is due to new information that is contained in the source of communication. Therefore, the goal of this paper is to develop a systematic methodology that allows us to summarize the information in major sources of communication and look at how changes in the presence of this information influence markets as signaled by changes in the yield curve. 
In our approach, we focus on identifying the main themes contained in the statements released by the Federal Open Market Committee (FOMC) directly after policy meetings and mapping changes in these themes back to the U.S. Treasury yield curve. 

We tackle this problem by borrowing tools from computer science. In the machine learning literature, natural language processing is a standard sub-field that has extensively developed a set of tools to work with text data. Within this sub-field, topic models are used to extract topics from a collection of documents. Only recently have these methods been extended to economics [as covered by Gentzkow et al. (2017)] and central bank communication [Hansen and McMahon (2016),  Hansen et al. (2017), Mazis and Tsekrekos (2017), Boukus and Rosenberg (2006), and Hendry (2012)]. However, of these applications, probabilistic methods are becoming a preferred method of topic modeling. As signaled by the name, these methods rely on drawing from probability distributions and assuming distributions of words given topics and topics given documents. One main issue with this is stability of the results. Often, you can re-run a probabilistic model and get different results each time, although different runs can be similar. This is clearly a disadvantage to using probabilistic topic models as changes in topic model results will ultimately influence later regression results.

Additional issues with probabilistic methods is their coherency and word composition. Often times, the topics generated may not be coherent and therefore cannot be given a structural interpretation, which is of importance when looking to identify the effects of specific topics of communication. These models tend to produce topics with words that are not typically associated with each other in a cohesive context. Furthermore, there can exist multiple topics with similar word compositions, which is repetitive information in our usage.

We look to address this issue by using a factorization topic model. This line of topic models looks to create matrix factors of an original (weighted) document-term matrix. The factors are usually determined by minimizing some objective function that measures the difference between the original matrix and the product of the two factors. Factorization methods have two advantages: (i) they are stable, meaning that the results will be the same each time you estimate the model, and (ii) they generally produce more coherent topics than probabilistic methods (O'Callaghan et al., 2015).

The final contribution of this paper is to introduce coherency measures into the economics literature for assessing the appropriateness of the topics. These coherency measures are metrics that look to determine how coherent a topic is based on the semantic relationships of the top words within the given topic. The advantage of using these measures is that they provide a natural way of choosing the number of topics to estimate within a model that does not fall subject to a researcher's a priori beliefs.

In summary, our contribution to the existing literature is three-fold: first, we use Non-negative Matrix Factorization (NMF) to uncover the topics; second, we use an automatic measure to select the number of topics; and, third, we relate these topics to several dimensions of the U.S. treasury yield curve.

\section{A Brief Literature Review}
\subsection{Central Bank Communication and Topic Modeling}
There exists a limited literature of applying topic models to central bank communication. This literature primarily focuses on the influence of communication on financial markets with a recent extension to macroeconomic outcomes. The methods cover both probabilistic and factorization methods used to summarize the sources of central bank communication.

The main method for probabilistically determining the topics across a corpus of central bank documents is Latent Dirichlet Allocation (LDA). Hansen, McMahon, and Prat (2017) use LDA to look at how transparency influences monetary policymakers' deliberations. They look to answer the question of whether language of the FOMC has changed since Transparency in 1993. More relevant to our study, Hansen and McMahon (2016) look at the effect of central bank communication on economic outcomes. The themes are captured using LDA on the FOMC statements, which identifies two main sources of information: current economic conditions and forward guidance. They then incorporate this into a factor-augmented VAR (FAVAR) with market and real economic variables and perform standard analysis to show that shocks to information on forward guidance is more important than communication about current economic conditions in determining variation of the economic variables.

One issue with the above approaches is that the topic models may suffer from the aforementioned issues with probabilistic models. Another issue would be that the number of topics is generally chosen arbitrarily. In the latter approach, they also use the FOMC statements and randomly chose to estimate a 15-topic model. The number of topics is an essential parameter and determines the topics that are found by the algorithm, therefore we would like to avoid \textit{a priori} researcher bias in choosing this parameter. In scientific research, a systematic approach to choosing the number of topics would be preferred, and this is a contribution we would like to make to the economics literature by using a measure of coherency.

Factorization methods have also been used in the central bank communication literature. Boukus and Rosenberg (2006) analyzes the FOMC Minutes using Latent Semantic Analysis (LSA), a singular value decomposition method. This finds topics in a similar manner as using principal components analysis to find a low dimensional representation of a highly dimensional data set. They find that there are significant financial market reactions to the specific themes. The reaction of treasury yields depends on the specific themes contained in the minutes. Mazis and Tsekrekos (2017) later take a similar approach and study the FOMC statements with LSA. They find that the themes have a significant effect on the change in the yield from medium- to long-term maturities using standard regression analysis. Relevant to this study, Hendry (2012) uses NMF to investigate what type of information from Bank of Canada communication statements or the market commentary based on these statements has a significant effect on the volatility or level of returns in a short-term interest rate market. They also find that different themes influence the market differently.

Mazis and Tsekrekos (2017) provide a framework that is similar to our study, however they choose a six-topic model based on a rule of thumb that they would take topics that explain five percent or more of the variation in the documents. Once again, we find that it is better to use a concrete measure that we would like to select the best model for us rather than depend on an arbitrary rule of thumb. Also in contrast, we would like to address the above issue as well as apply Non-negative Matrix Factorization (NMF) as in Hendry (2012) but to the FOMC statements.

\subsection{Central Bank Communication and Financial Markets}
As mentioned earlier, central bank communication is capable of influencing expectations of future short-term rates, which influence long-term rates and financial market prices. The application of the methodology presented in this paper is primarily on the influence of central bank communication on financial markets and we will present the relevant ideas and findings in the literature in this section.

The previous literature has focused on addressing two main issues in identifying the impact of central bank communication on financial markets. First, there exists the general question of how to identify communication and how to define an `event.' Second, the literature looks to address the how to extract the intention or objective behind a policy statement in order to assess whether the statement was successful.

To address these issues, a few studies have simplified their questions to look at general `differences' on days of policy statement releases. The most common simplification is to look for significant differences in the volatility of financial variables on days of statement releases [Kohn and Sack (2004), Connolly and Kohler (2004), and Reeves and Sawicki (2007)]. These studies focus on the influence of communication on returns of financial assets, and hypothesize that on days of statement releases there should be higher volatility. The largest advantage of this simplification is that it is not necessary to analyze the direction of impact. Ultimately, this allows the researcher to avoid the challenging question of how to identify the polarity of each statement, which is essential since we would like to know, for example, whether stock prices fall due to negative information or vice versa. This is an appropriate method to study whether central bank communication creates news and not an attempt to predict which direction markets may move in. Therefore, we find that volatility is an appropriate measure that we would like to capture in our study as well.

Kohn and Sack (2004) show that the release of the FOMC policy statements significantly impact the volatility of various asset prices. They conclude that this provides evidence that there exists relevant information within these statements that the markets react to. Specifically, they find that statements affect interest rates over short-to-medium horizons. Reeves and Sawicki (2007) find similar results with the same approach but looking at communication by the Bank of England. In particular, the Monetary Policy Committee minutes and the \textit{Inflation Reports} significantly impact financial markets.

The literature has addressed issues with focusing on volatility. The first issue is that financial market volatility can be determined by many factors other than central bank communication, such as additional news and the current state of the economy (Reeves and Sawicki, 2007). Additionally, communication can be endogenously determined by economic conditions itself. However, as stated by Blinder et al. (2008), endogeneity is less of a problem when the release dates of the major communication are known in advance. This is the case for the FOMC policy statements as the dates and times are posted online months in advance.

Another strand of this literature looked to address the previous issues of assessing whether communication had its intended effect and predicting the direction the market will move in. These studies attempted to quantify communication to look at the direction and magnitude of its impact on financial markets. These studies code directional indications of the statements, such as a positive sign for hawkish statements, negative for dovish, and zero for neutral statements [Jansen and De Haan (2005), Ehrmann and Fratzscher (2007)]. Other studies extend this method to a grid of values that look to make a more detailed suggestion of magnitude, such as using a scale of -2 to 2 [Rosa and Verga (2007), Musard-Gies (2006)]. 

Ehrmann and Fratzscher (2007) find that statements tend to move financial markets in their intended direction. Statements that suggest tightening lead to interest rate increases, while those that suggest easing lead to lower interest rates. Musard-Gies (2006) finds similar results for the European Central Bank (ECB), showing that the short end of the yield curve reacts more than the long end. However, in the U.S., statements related to economic outlook are found to mainly influence the medium and long end of the yield curve. Other studies are conducted analyzing speeches within different European countries [Rozkrut et al. (2007), Andersson et al. (2006)].

However, all of these methods have fallen subject to the author's interpretation of the statement. As mentioned in Blinder et al. (2008), classifying these documents is a subjective task and also faces the possibility of misclassifications. This risk can be remedied but not eliminated by using methods of content analysis combined with an approach such as having independent classifications by multiple researchers (Berger et al., 2006).

Additionally, the indicators are \textit{ex post} measures and therefore may not reflect the actual perception of financial markets at the time of release. The actual perception of financial markets are likely to be determined by the expectations of monetary policy and the current understanding of the state of the economy at the time of the release of the statement. Markets should solely react to the unexpected component of communication and that may be different from what may be suggested by ex-post measures of the direction. For example, hypothetically, markets may expect the Federal Reserve to introduce a large increase in the federal funds rate, however if the federal reserve increases the rate by a much smaller amount, then markets will react to this unexpected difference.

A seminal study that addressed this issue is by Gürkaynak et al. (2005) which used principal components analysis to identify two factors that describe asset price movements around the release of FOMC policy statements. They identify a communication effect by attributing it to the component that is orthogonal to the federal funds rate. This factor is found to affect interest rates across the entire yield curve, but appears to be most sizable at the long end of the yield curve. Recently, using a similar methodology for studying the ECB, Leombroni et al. (2017) find that communication shocks have the most pronounced impact at intermediate maturities, generating a humped-shaped response in the term structure.

\section{Data}
The main data set combines multiple sources of data available at a daily frequency over the period of May 10, 1999-October 20, 2017. These include the full set of statements released directly after the regularly scheduled FOMC meetings and two sources for U.S. Treasury yields.

\subsection{FOMC Statements}
We focus on information communicated by the Federal Reserve. To measure the sources of information in communication by the Federal Reserve, the FOMC statements are scraped from the Board of Governors of the Federal Reserve System website\footnote{https://www.federalreserve.gov}. This provides a time series of statements over the period of May 10, 1999 through October 20, 2017. Structurally, they are released on the final day of the FOMC monetary policy meetings at 2:00 pm. These statements provide the first opportunity for monetary policy decisions to be communicated to the public, as well as relay information related to the state of the economy and financial system.

Since 1981, the FOMC has eight regularly scheduled meetings per year. This excludes special meetings and telephone conferences that are held under extenuating circumstances. From the beginning of these regular scheduled meetings there were no information releases until February 1994. From then up until mid-1999, the FOMC issued statements after policy meetings only when there were decided changes in the federal funds rate. Finally, in May 1999, the FOMC decided to release a statement after every meeting.

\begin{figure}[!h]
\begin{center}
\caption{Raw and Preprocessed Statement Lengths}
\includegraphics[scale=0.85]{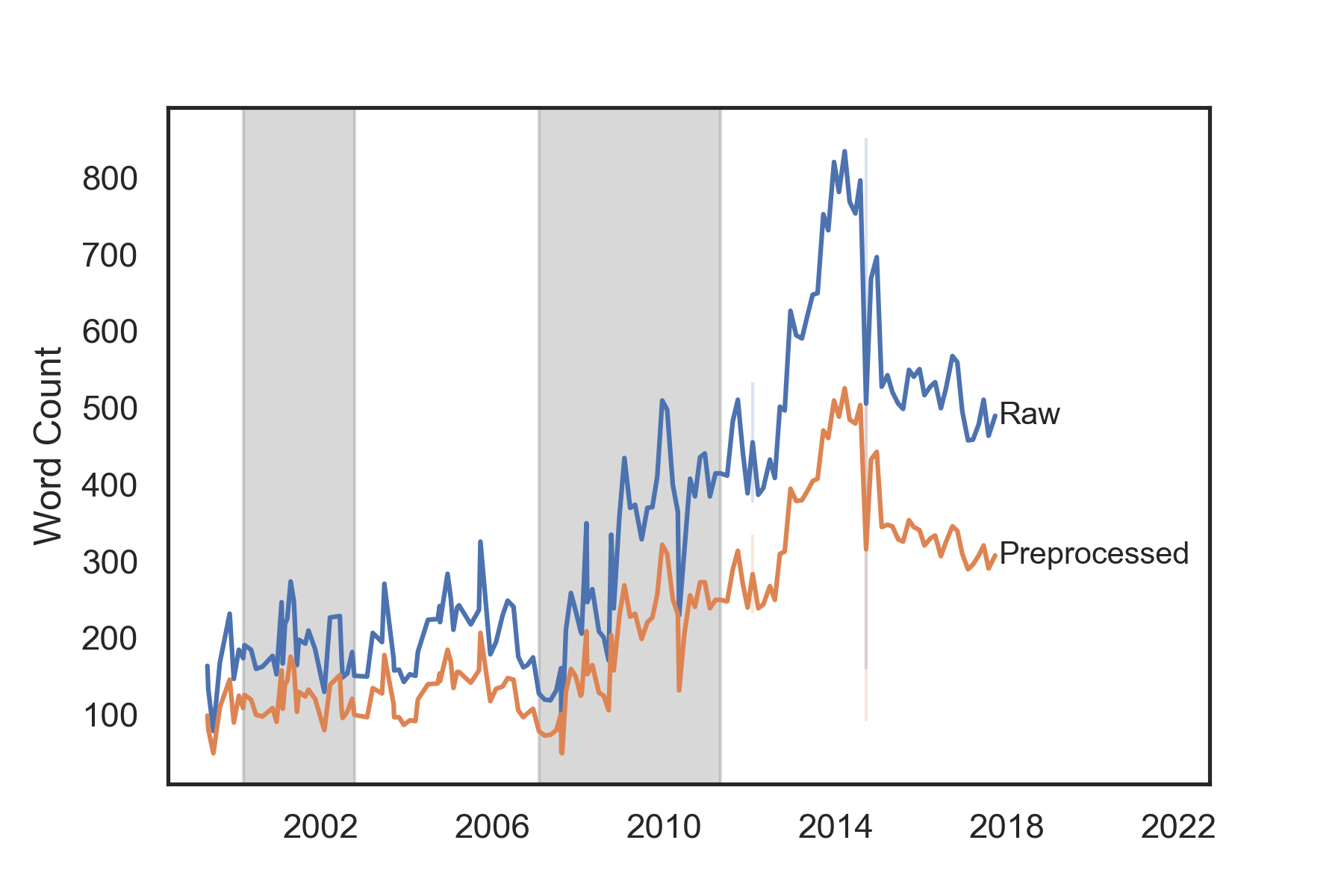}
\label{fig:doclength}
\floatfoot{Note: This figure shows the counts of terms over time for the raw and preprocessed text. These are counts after removal of voting outcomes in the statements.} 
\end{center}
\end{figure}

It is important to address some of the issues with identifying themes in the FOMC statements. First, there has been an upward trend in the length of the statements. This is documented in both the raw and the preprocessed document word count. This can be seen in Figure \ref{fig:doclength}. This trend is likely to be driven by the focus on increased transparency by central banks in recent years, especially after the Global Financial Crisis of 2007, contributing to an emphasis on effective communication. Second, as reported in Rosa (2011, 2013), the initial releases of the statements were a trial phase and they did not settle on a final form until around mid-2003. Third, the statements are released under various chairmen of the Federal Reserve and may represent language related to that Chairman. Statements between May 1999 and January 2006 fall under Alan Greenspan; February 2006 to January 2014 fall under Ben Bernanke; and, February 2014 to October 2017 fall under Janet Yellen. Finally, we should note that the time span contains the early 2000's recession (2000-2002) attributed to the Dot-Com Boom, as well as the well-known Global Financial Crisis spanning 2007-2011. These events could have endogenously determined the language in the statements, including word usage and the generation of new topics.

\subsection{U.S. Treasury Yields}
The first source of data for the U.S. Treasury yields is published by Gürkaynak et al. (2007). They provide U.S. Treasury yields from 1961 to the present on a daily frequency. The quotes provided in their dataset are derived from two different sources. From June 14, 1961 to the end of November 1987, the quotes on Treasury securities are collected from the Center for Research in Security Prices (CRSP), which provides end-of-day quotes on all outstanding Treasury securities. From December 1987, the quotes are collected from the Federal Reserve Bank of New York, which is constructed from several sources of market information (Gürkaynak et al., 2007). Since we are specifically analyzing the FOMC statements from May 1999 and onward, our dataset will consist of yields from the latter source.

The main advantage of using this data set for our analysis is that it provides yields for maturities from 1 to 30 years in one-year intervals. This provides a rich source of information for estimation of the daily yield curve factors later in the analysis.

However, in estimation of the yield curve factors, we would like to include shorter term maturity yields. It is well-known that the short end of the yield curve is closely tied to short-term interest rates, which are determined by monetary policy decisions. Therefore, we augment the Gürkaynak et al. (2007) data with the 3- and 6-month U.S. Treasury yields published directly by the U.S. Department of the Treasury\footnote{https://www.treasury.gov/resource-center/data-chart-center/interest-rates/Pages/TextView.aspx?data=yield}. Unfortunately, the U.S. Department of the Treasury does not publish a substantial range of maturities to use it as the main source of data\footnote{The U.S. Department of the Treasury publishes the 1 month, 3 month, 6 month, 1 year, 2 year, 3 year, 5 year, 7 year, 10 year, 20 year, and 30 year daily Treasury yield curve rates.}. This is disadvantageous because it provides less information for the estimation of the yield curve. Therefore, we prefer a combination of both sources to effectively get an empirical representation of the yield curve.

\subsection{Macroeconomic and Financial Market Variables}
Our dataset includes variables collected from the FRED Economic Database of the Federal Reserve Bank of St. Louis that relate to economic and financial market conditions. In our models, we look to control for financial market stress, economic conditions, and expectations about the business cycle and future economic conditions. To suit our analysis, we need variables that control for these at a daily frequency. As in Mazis and Tsekrekos (2017), we include the term spread, credit spread, and financial market volatility controls. 

The term spread we use is the 10-year constant maturity minus the 3-month Treasury constant maturity. As commonly mentioned, the term spread looks to control for economic conditions and short-term prospects.

The credit spread we use is the ICR BofAML US Corporate BBB Option-Adjusted Spread (OAS). The credit spread acts as a proxy for expectations about the business cycle and future changes in the economy. In general, this OAS is the calculated spread between a computed OAS index of all corporate bonds with BBB rating and a spot Treasury curve. The OAS index is constructed by using the OAS and weighting it by market capitalization.

Finally, as a control for financial market volatility, we use the CBOE Volatility Index (VIX), which is a standard variable to control for financial market stress. Officially, the VIX measures the market expectation of near term volatility conveyed by stock index option prices.

\section{Empirical Methodology}
Our approach to identifying the effects of information within the FOMC statements on the yield curve is four-fold. First, we perform text preprocessing and feature extraction to transform the statements into a data set that can be used to determine the topics. Second, we use a topic modeling algorithm, namely Non-negative Matrix Factorization (NMF), to determine the main topics present in the collection of statements and we compare the quality of these topics to a competing method, Latent Dirichelet Allocation (LDA), by using coherency measures. Third, we use the U.S. Treasury yields data and estimate the Diebold and Li (2006) model, extracting factors corresponding to the level, slope, and curvature of the yield curve. Finally, we use regression analysis to determine if these topics have a significant impact on fluctuations in the estimated factors.

\subsection{Text Preprocessing and Feature Extraction}

The purpose of performing text preprocessing is to strip the collection of documents of irrelevant information. This process is dependent on the specific application of topic modelling. Our first steps to preprocessing the collection of documents are to remove stopwords and names of FOMC members. Stopwords are common words that do not offer any textual significance, but is present simply for grammatical and structural purposes. Common stopwords are 'the', 'and', 'but', 'if', etc.  We remove the voting procedures and outcomes that usually occur at the end of a statement, because this does not relay any information related to the main topics that are covered in the statements. 

\begin{figure}[!h]
\begin{center}
\caption{Text Preprocessing Example}
\includegraphics[scale=0.37, trim={0 2.5cm 0 2.5cm},clip]{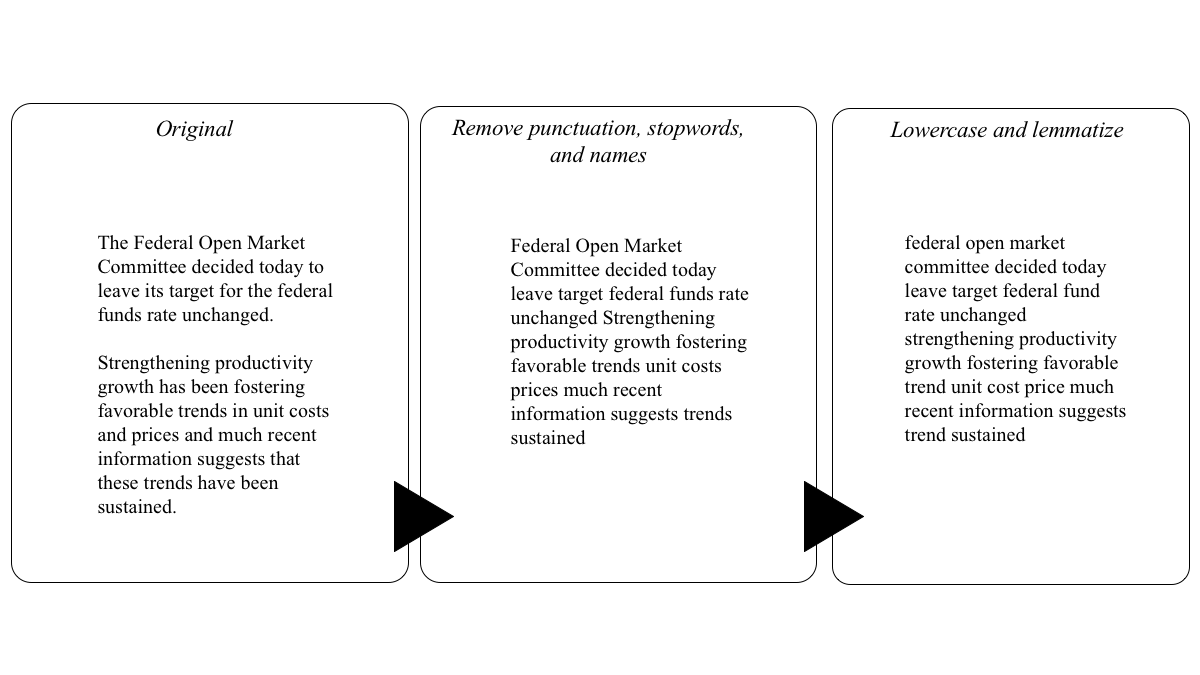}
\floatfoot{Note: Excerpt from FOMC statement, May 10, 1999.} 
\label{fig:txt}
\end{center}
\end{figure}

Our next steps are to lower case words and lemmatize the words as processes of standardizing the text. Lowercasing each word in the documents allows us to ignore capitalization so that ``Inflation" and ``inflation" will both be treated the same within the modelling process. The goal of lemmatization is to transform plural forms of words to their singular form and also transform past-tense verbs to their present-tense form (Mazis and Tsekrekos, 2017). This process properly standardizes any morphological affixes of given words. For example the words ``economic", ``economy", and ``economical" will all be returned as ``economy." The text preprocessing steps are demonstrated in Figure \ref{fig:txt}.

Next, we tokenize all of the documents, which splits each statement into a collection of its individual words. We then take all of the tokenized documents and transform them into a Bag-of-Words model. The Bag-of-Words representation yields an $n\times m$ matrix with $n$ documents on the rows and $m$ unique terms on the columns. The $c_{nm}$ entry of the matrix yields the count of times term $m$ appears in document $n$. Fitting the collection of documents into this model is advantageous as it gives us numerical measures that we can then feed into a machine learning algorithm.

As a final step to text preprocessing, we perform term weighting, which improves the usefulness of the document-term matrix that yields from the bag-of-words representation. We use the term frequency-inverse document frequency (TF-IDF) weighting scheme, which gives higher weights to more ``important" terms. Term frequency ($tf$) relates to the number of times a term appears in a single document. As a standard measure, $tf(n,m) = 1 + log(c_{nm})$. Inverse document frequency ($idf$) refers to a function of the total number of distinct documents the word appears in, which penalizes terms that appear in a large number of documents. This measure is summarized in the following equation:
\begin{equation}
w(n, m) = tf(n, m) \times (log(\dfrac{D}{df(m)}+1),
\end{equation}
where $D$ is the total number of documents.

We can see that when using the TF-IDF weighting scheme terms that appear frequently in a low number of documents will have a high weight. This measure attempts to capture the fact that these terms are important to that specific document and are essential to the topics presented within the document. Therefore, this measure provides an intuitive representation of the document-term matrix that allows us to find important topics while placing an emphasis on the document-defining terms within the corpus. An additional benefit of using TF-IDF weighting is that it naturally penalizes domain-specific stopwords that are likely to appear in many documents.

\subsection{Non-Negative Matrix Factorization}
\subsubsection{Algorithm}
\begin{figure}[h]
\begin{center}
\caption{Non-Negative Matrix Factorization (NMF) for Topic Modelling}
\includegraphics[scale=0.30, trim={0 2.5cm 0 2.5cm},clip]{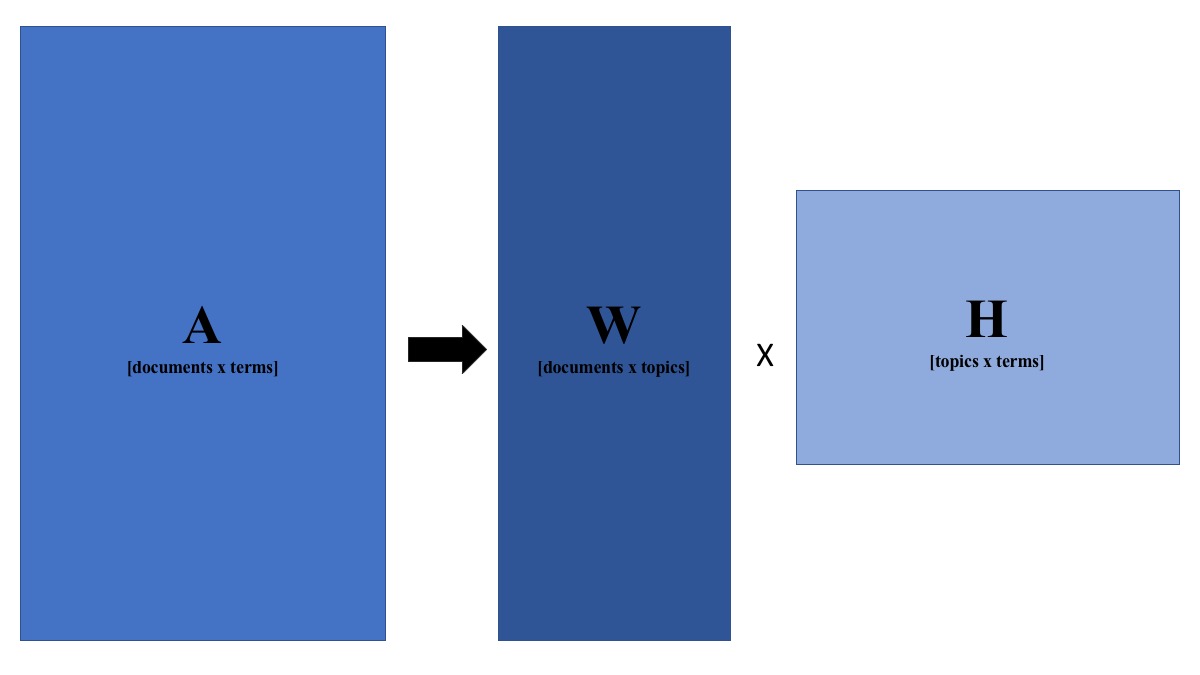}
\floatfoot{Note: NMF decomposes a non-negative matrix $A$ into two factors $W$ and $H$ by attempting to minimize the Euclidean distance between $A$ and the product of $W$ and $H$.} 
\label{fig:NMF}
\end{center}
\end{figure}

Using the TF-IDF weighted matrix, we then use the non-negative matrix factorization algorithm by Lee and Seung (1999) for topic modeling. The objective of this algorithm is to decompose a non-negative matrix $A$ into two matrix factors, $W$ and $H$. This can be visualized in Figure \ref{fig:NMF}. The $W$ matrix is $n \times k$ with each entry $w_{nk}$ representing the weight of topic $k$ in document $n$. The $H$ matrix has dimension $k \times m$ with each entry $h_{km}$ representing the weight of term $m$ in topic $k$. The algorithm uses a local-EM style optimization procedure to minimize the following objective function:
\begin{equation}
\dfrac{1}{2}||\textbf{A} - \textbf{WH}||^{2}_{F}=\dfrac{1}{2} \sum^{n}_{i=1}\sum^{m}_{j=1}(A_{ij} - (WH)_{ij})^{2}.
\end{equation}
We prefer to use NMF based on findings by O'Callaghan et al. (2015) which show that factorization methods to topic modeling outperform probabilistic methods in extracting coherent topics from documents associated with niche fields. Personally, we believe that economics should be considered a niche field given its use of specific toolsets, ideas, and jargon. We believe this makes NMF more suitable for text mining applications in economics. A brief comparison between NMF and its competing probabilistic model Latent Dirichlet Allocation is shown in section 4.3.

\subsubsection{Model Selection}
When performing topic modeling, $k$ (i.e the number of topics) is a parameter that must be chosen by the researcher. The number of topics can either be chosen subjectively, based on previous knowledge, or selected automatically based on various measures of semantic relationships among the words. 

We believe that the most robust way of determining these topics is to use the latter approach. A common measure that is present in the machine learning literature, which is currently neglected in the economics literature, is the coherency measure. Coherency measures attempt to measure the sensibleness of topics yield by a model. This offers a scientific benefit of introducing less bias in choosing $k$, avoiding identifying topics matching those based on the researcher's \textit{a priori} expectations.  

Mimno et al. (2011) proposed to use an asymmetrical coherence measure between top word pairs. With mean log conditional probability coherence (LCP), also referred to as `UMass Coherence', word probabilities are estimated based on document frequencies of the original documents used for learning the topics (Röder et al., 2015). The original metric for computing LCP presented by Mimno et al. (2011) was
\begin{equation} 
TC-LCP^{(t)} = \sum_{i=2}^{N}\sum_{j=1}^{i-1} log\dfrac{D(w_{i}^{(t)}, w_{j}^{(t)}) + 1}{D(w_{j}^{(t)})},
\end{equation}
where $D(w_{i}^{(t)}, w_{j}^{(t)})$ represents the co-document frequency of the words  $w_{i}$ and $w_{j}$ in topic $t$. $D(w_{j}^{(t)})$ is the document frequency of $w_{j}$ in topic $t$. In general, this measure plans to capture the coherence of each topic by summing the co-occurence of each word with other words in the topic weighted by the total number of times the word occurs. The summation accounts for the ordering of the words from most probable (high weighted) to least probable (lowest weighted). Therefore, high coherency will found for those topics that have high co-frequency of the most probable words in the topic throughout all of the documents, but it will penalize those topics that contain high-frequency words.

As Röder et al. (2015) show, this is equivalent to the empirical conditional log probability of $w_{j}$ given $w_{i}$. They also provide a slightly modified version of the measure that adds a constant outside this probability that takes an average of the summed probabilities based on combinatorics:
\begin{equation}
TC-LCP*^{(t)} = \dfrac{2}{N (N-1)} \sum_{i=2}^{N} \sum_{j=1}^{i-1} log\dfrac{P(w_{i}^{(t)}, w_{j}^{(t)}) + \epsilon}{P(w_{j}^{(t)})}, \footnote{The original paper by Mimno et al. (2011) assumed $\epsilon = 1$, but Stevens et al. (2012) found that UMass coherency performs better when $\epsilon$ is rather small.},
\end{equation}
where $\dfrac{P(w_{i}^{(t)}, w_{j}^{(t)})}{P(w_{j}^{(t)})}$ is equivalent to $P(w_{i} | w_{j})$. Taking the log of this gives the log conditional probability and smoothing is introduced by adding $\epsilon$ to the numerator to avoid taking the logarithm of zero. In general, this measure evaluates co-occurences of terms in a way that it pays attention to whether a more probable (higher weighted) word within a specific topic can predict a less probable (lower weighted) one. If this conditional probability is high with respect to the amount of times the more probable word occurs in the corpus, then a high coherency will be achieved.

\begin{figure}[h]
\begin{center}
\caption{NMF Coherency}
\includegraphics[scale=0.85]{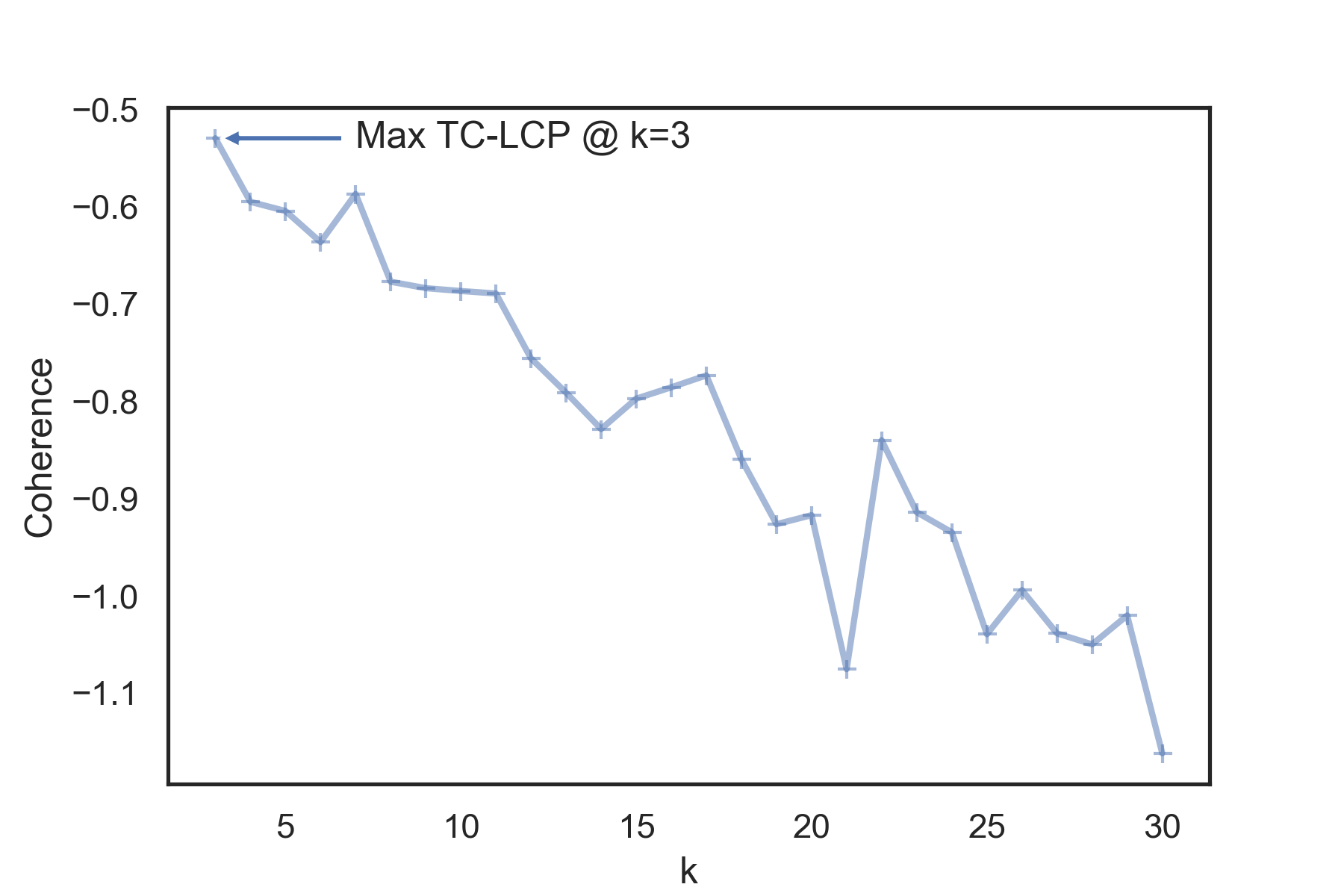}
\floatfoot{Note: Average TC-LCP coherency measured on range of NMF models with $k \in [3, 30]$ and unigrams.} 
\label{fig:UMassNMF} 
\end{center}
\end{figure}

We calculate the modified UMass coherency for a range of NMF models with $k \in [3,30]$. We evaluate the coherency of these topics using the top 15 words ($N = 15$) in each topic. After calculating coherency across the range of models, our maximum coherency is achieved where $k=3$. The results are shown in Figure \ref{fig:UMassNMF}.

We compare this optimal NMF model with $k=3$ to additional models to assess whether we believe the coherency measure is indeed finding a an optimal model. The models we compare this to are the NMF with $k=4$ and $k=5$. These models do not have a significant coherence difference and thus lets us compare between slightly similar models based on coherence. The results of this analysis are reported in Table \ref{table:NMFTopics}. From looking at the NMF model with $k=4$, we can see that the first three topics in this model are comparable to the optimal model. There is a large amount of word sharing in this model with all of the words in the first topic being the same as the words in the first topic of the optimal model albeit with minor ranking differences. The second topic is 53 percent composed of words that are contained in the second topic of the optimal model. Topic three contains 83 percent of the words that are in topic three of the optimal model.

\begin{landscape}
\begin{table}[!ht]
\centering
\begin{adjustbox}{width=\textwidth, height=.25\textheight, center=\textwidth}
\caption{Comparing Topics of NMF Models}\label{table:NMFTopics}
\begin{tabular}{c c c c c c c c c c c c c}
\hline
& \multicolumn{12}{c}{}\\
 & \multicolumn{3}{c}{$k=3$} & \multicolumn{4}{c}{$k=4$} & \multicolumn{5}{c}{$k=5$} \\
  & \multicolumn{3}{c}{$\overbrace{\qquad \qquad \qquad \qquad \qquad \qquad \quad}$} & \multicolumn{4}{c}{$\overbrace{\qquad \qquad \qquad \qquad \qquad \qquad \qquad \qquad \qquad \quad}$} & \multicolumn{5}{c}{$\overbrace{\qquad \qquad \qquad \qquad \qquad \qquad \qquad \qquad \qquad \qquad \qquad \qquad \quad}$} \\
 & \multicolumn{12}{c}{} \\
 \textit{Rank} & 1 & 2 & 3 & 1 & 2 & 3 & 4 & 1 & 2 & 3 & 4 & 5\\
\hline
1 & inflation & board & \textbf{security} & inflation & board & \textbf{security} & jr & \textbf{inflation} & board & \textbf{security} & jr & \textbf{inflation}\\
2 & longer & growth & credit & longer & approved & credit & \textbf{growth} & longer & action & level & monetary & growth \\
3 & labor & approved & financial & labor & basis & recovery & belief & labor & approved & recovery & belief & moderate\\
4 & term & action & \textbf{reserve} & term & action & level & monetary & term & basis & \textbf{reserve} & sustainable & commodity\\
5 & \textbf{condition} & basis & recovery & condition & point & \textbf{reserve} & policy & \textbf{condition} & point & credit & growth & risk\\
6 & policy & point & level & policy & \textbf{growth} & financial & available & \textbf{security} & discount & subdued & available & evolution\\
7 & price & discount & exceptionally & \textbf{security} & discount & exceptionally & sustainable & employment & \textbf{reserve} & exceptionally & \textbf{policy} & core\\
8 & \textbf{security} & \textbf{reserve} & purchase & fund & \textbf{reserve} & subdued & circumstance & \textbf{percent} & bank & purchase & bies & reflecting\\
9 & fund & sustainable & continue & price & bank & purchase & roger & \textbf{policy} & request & low & ferguson & recent\\
10 & run & productivity & promote & percent & request & continue & susan & range & related & financial & susan & needed\\
11 & employment & belief & subdued & employment & related & low & ferguson & run & \textbf{percent} & continue & roger & effect\\
12 & pace & risk & facility & range & submitted & promote & bies & agency & submitted & facility & productivity & quarter\\
13 & percent & demand & billion & run & director & facility & gramlich & pace & director & resource & gramlich & likely\\
14 & range & price & \textbf{condition} & pace & governor & billion & edward & price & governor & \textbf{condition} & edward & energy\\
15 & agency & available & housing & agency & taking & resource & rice & fund & taking & billion & rice & implied\\
\hline
\hline
\end{tabular}
\end{adjustbox}
\begin{tablenotes}
 \small
 \item \textit{Notes:} This table shows the top 15 words for topics given by three different NMF models. We estimate NMF models for $k \in [3,5]$. Each column represents the respective topic for the given model in brackets, where these topics are given by the columns of the $W$ matrix associated with that model. The top words are ranked based on their weight in the respective row of the $H$ matrix from greatest to least. Words in \textbf{bold} are words that overlap between topics.
\end{tablenotes}
\end{table}
\end{landscape}

A main difference of the $k=4$ model is the additional topic. This topic is mixed with terms related to economic conditions such as `growth', terms related to policy like `monetary' and `policy', and names such as `ferguson' and `edward'. The mix of these different terms and concepts within a single topic are likely a reason why average coherency of the model is lower than the optimal model. Additionally, it would be difficult to give a structural interpretation of to this topic that would be relevant for economic analysis. Therefore, this model would be least preferred compared to the optimal model chosen based on the coherency measure.

Now comparing the NMF model with $k=5$ to the optimal model, we can see that the first three topics also coincide with the three topics in the optimal model. The first topic consists of the same words as the first topic in the optimal model, but with ranking differences. Approximately 47 percent of the second topic consists of words in the second topic of the optimal model. The third topic is 86 percent made up of words in the third topic of the optimal model. We can conclude that the first three topic are robust to adding additional topics.

The additional two topics in the NMF model with $k=5$ are less coherent and interpretable. The fourth topic is similar to the fourth topic of the NMF model with $k=4$. It contains words related to economic conditions, policy, and names. The fifth topic contains words related to economic conditions such as `inflation', `growth', and `risk', but also includes words such as `likely', `needed', `recent', and 'implied' which are not necessarily terms related to conditions but are expected to be used with other important economic concepts. However, these terms can be used in any context and is not guaranteed to be consistently used with the other words in this topic. This may yield a low coherency value, which returns a low average coherency. It is also difficult to attribute a structural interpretation to this topic, which would not be preferable for use in econometric modeling. Therefore, the optimal model is preferred over this model as well.

\subsubsection{Results}
After estimating our best model based on the coherency measure above, we can interpret the FOMC statements as relaying information related to three topics. The topics are associated with a set of words that all come with a weight of that word within the respective topic. Words with higher weights have the most importance within that topic and therefore defines the information in the topic.

\begin{figure}[H]
\begin{center}
\caption{Top 15 words of Theme 1, NMF}
\includegraphics[scale=0.60]{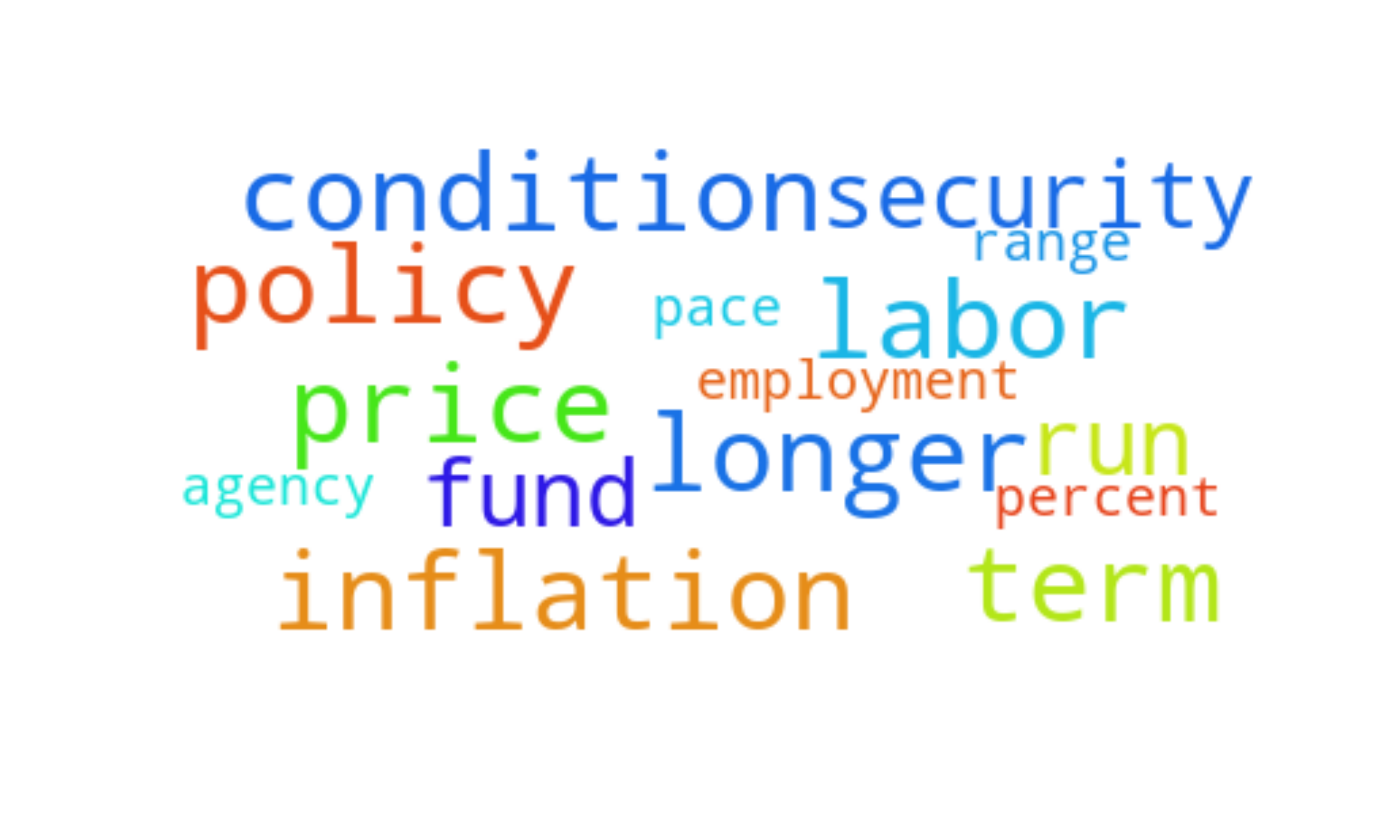}
\label{fig:nmf1}
\end{center}
\end{figure}

\begin{figure}[!h]
\begin{center}
\caption{Top 15 words of Theme 2, NMF}
\includegraphics[scale=0.60]{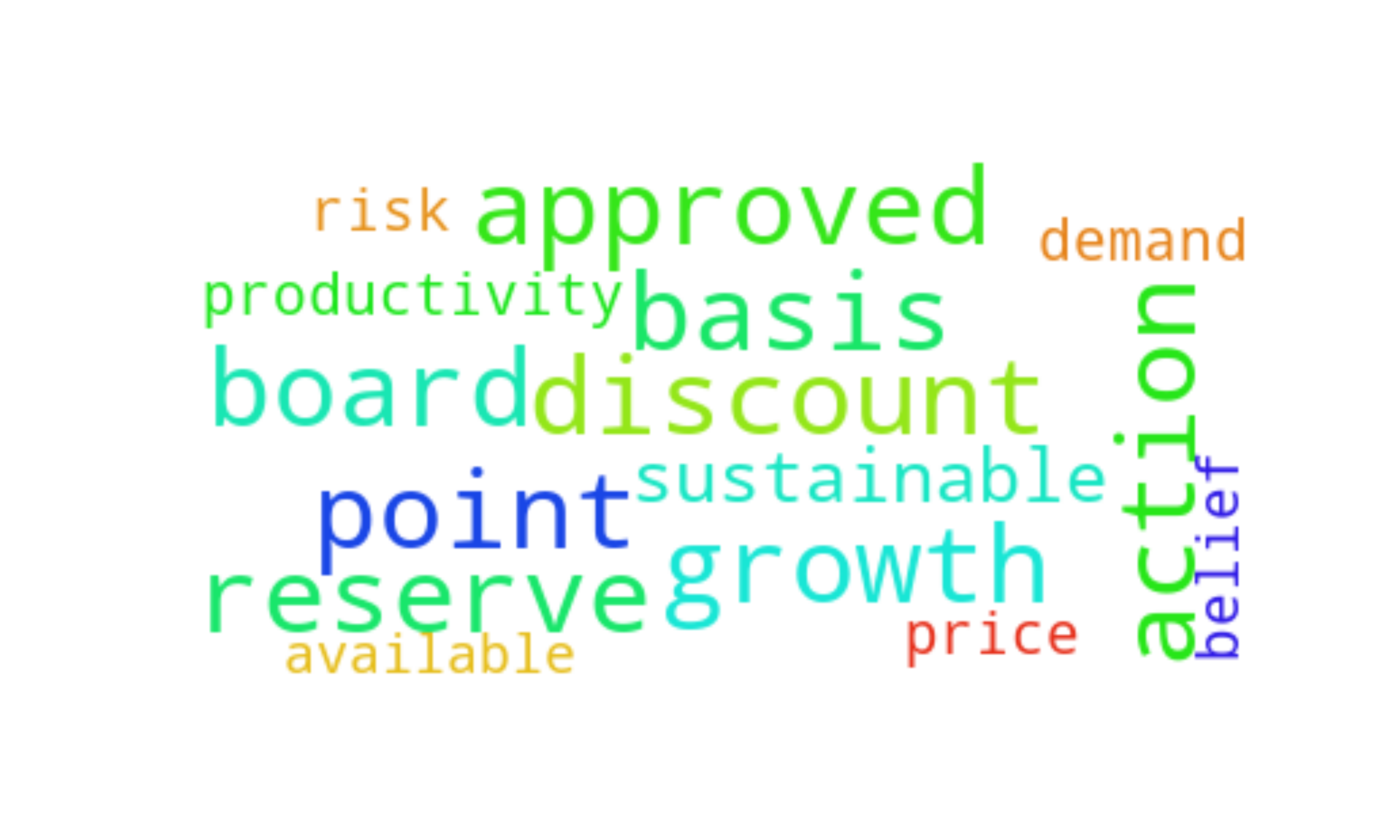}
\label{fig:nmf2}
\end{center}
\end{figure}

\begin{figure}[!h]
\begin{center}
\caption{Top 15 words of Theme 3, NMF}
\includegraphics[scale=0.60]{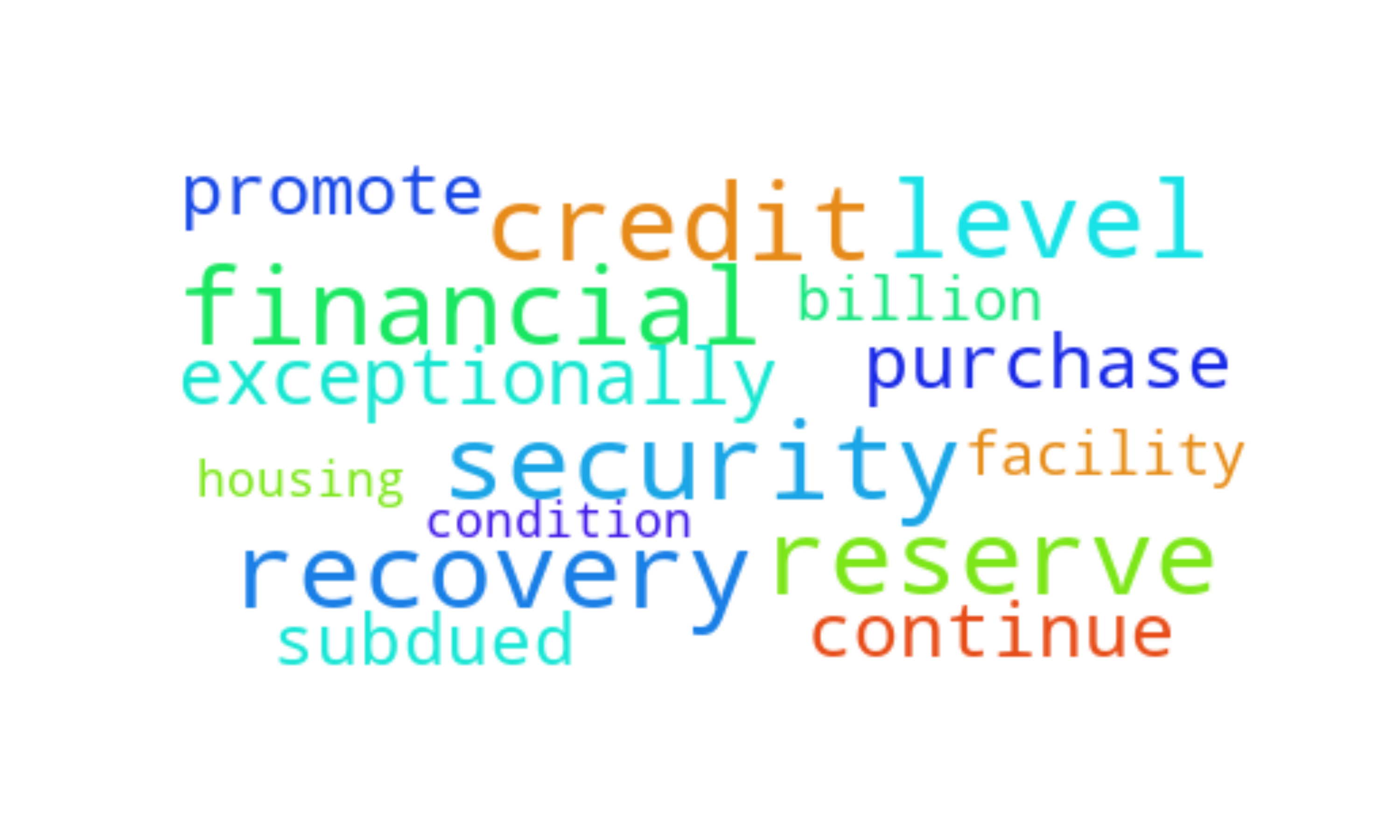}
\label{fig:nmf3}
\end{center}
\end{figure}

Naturally, from this set-up, we would like to look at the words with the highest weights to ascertain an interpretation to these topics. Therefore, we look at the top 15 words of the topics used in the evaluation of coherency and assume a 'topic label' that describes the theme of that topic. In general, we find that the topics can be decomposed in to (i) information related to the mandates, (ii) information related to monetary policy tools, and (iii) information related to financial markets. The top 15 words for each theme/topic are reported in Figures \ref{fig:nmf1}-\ref{fig:nmf3}.

In topic 1, henceforth referred to as `Theme 1,' we see the presence of words such as `employment', `labor', `inflation', and `security'. We believe this topic relates to the objectives of the Federal Reserve, which follow from the dual mandate of price stability and maximum employment.

In theme 2, we see words such as `basis', `action', `discount', and `point.' We believe that the presence of these words relate to the monetary policy tools. Words such as `growth', `demand', and `risk' are words related to targets that the central bank may use monetary policy to influence. Ultimately, we attribute this topic to be a monetary policy topic.

Finally, in theme 3, we can see words such as `financial', `security', `housing', and `recovery'. These words are characteristic of information covering financial markets and the financial crisis of 2007-2011. Words such as 'continue' likely relate to monetary policy around the financial crisis, around which the Federal Reserve often stated that it will continue to keep interest rates at the zero lower bound. In general, we refer to this topic as a `financial topic.'

\begin{figure}[!h]
\begin{center}
\caption{Theme 1 ('Economic Conditions and Mandates') Weights, NMF}
\includegraphics[scale=0.80]{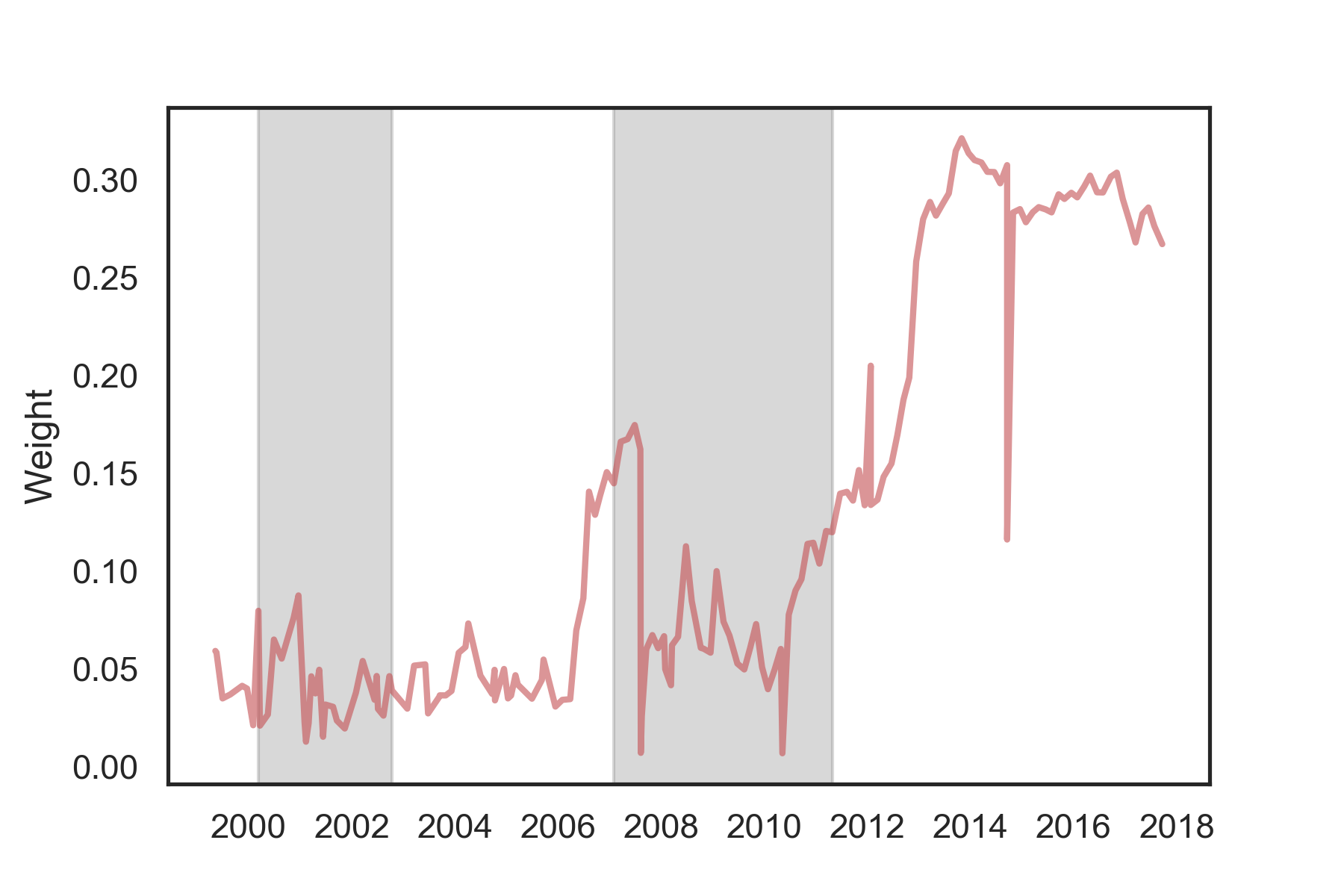}
\label{fig:nmfweights1}
\floatfoot{Notes: The weights of topic one over each statement is plotted as a time series. These weights correspond to the first column of the $W$ matrix derived from the NMF algorithm. These weights are estimated from the NMF model with $k = 3$ and only unigrams considered. The shaded regions represent the crisis periods present in the sample. From left to right, the first shaded region represents the crisis after the bursting of the Dot Com bubble. The second shaded region represents the global financial crisis.}
\end{center}
\end{figure}

\begin{figure}[!h]
\begin{center}
\caption{Theme 2 ('Monetary Policy Tools and Intermediate Targets') Weights, NMF}
\includegraphics[scale=0.80]{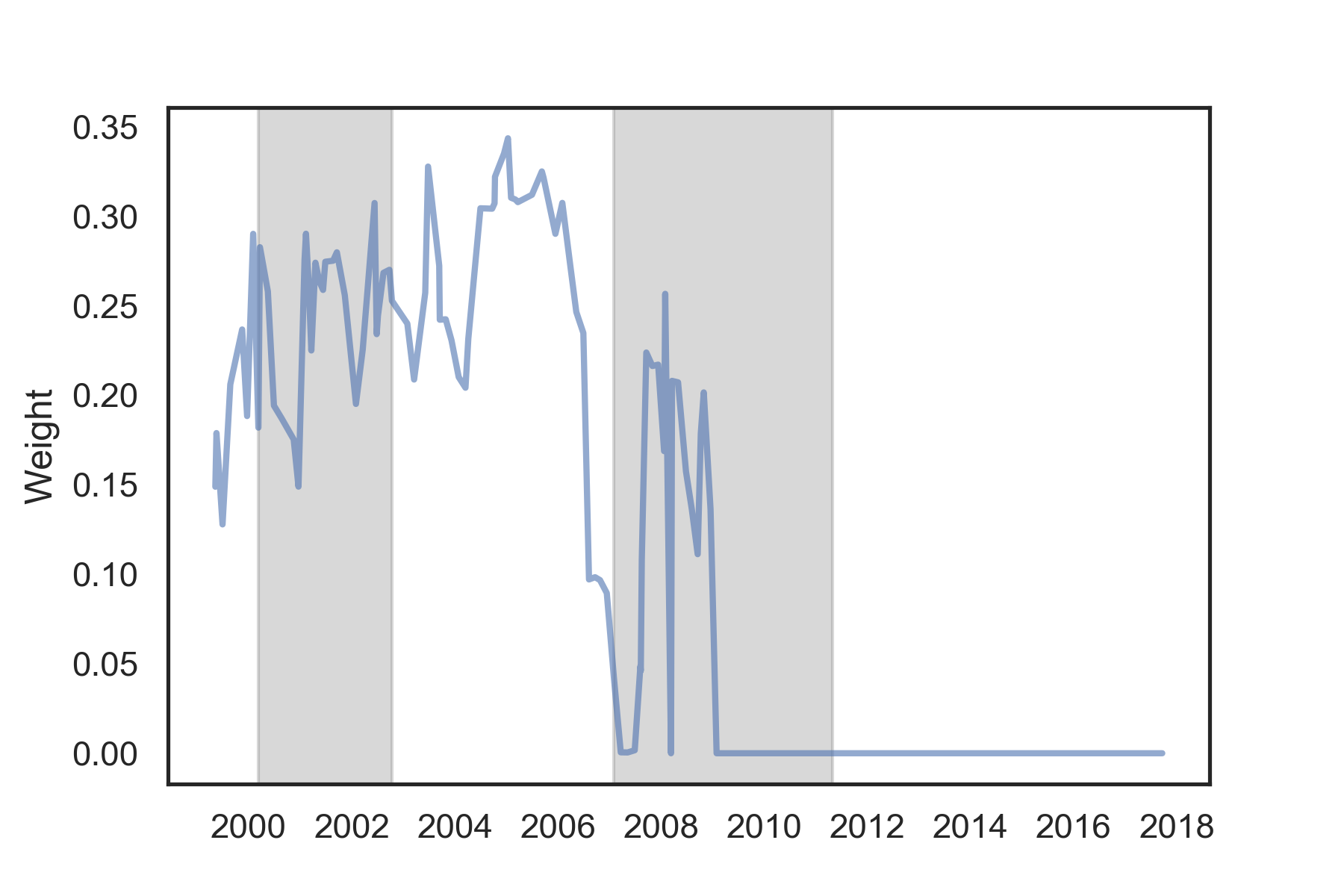}
\label{fig:nmfweights2}
\floatfoot{Notes: The weights of topic two over each statement is plotted as a time series. These weights correspond to the second column of the $W$ matrix derived from the NMF algorithm. These weights are estimated from the NMF model with $k = 3$ and only unigrams considered. The shaded regions represent the crisis periods present in the sample. From left to right, the first shaded region represents the crisis after the bursting of the Dot Com bubble. The second shaded region represents the global financial crisis.}
\end{center}
\end{figure}

\begin{figure}[!h]
\begin{center}
\caption{Theme 3 ('Financial Markets and Financial Crisis') Weights, NMF}
\includegraphics[scale=0.80]{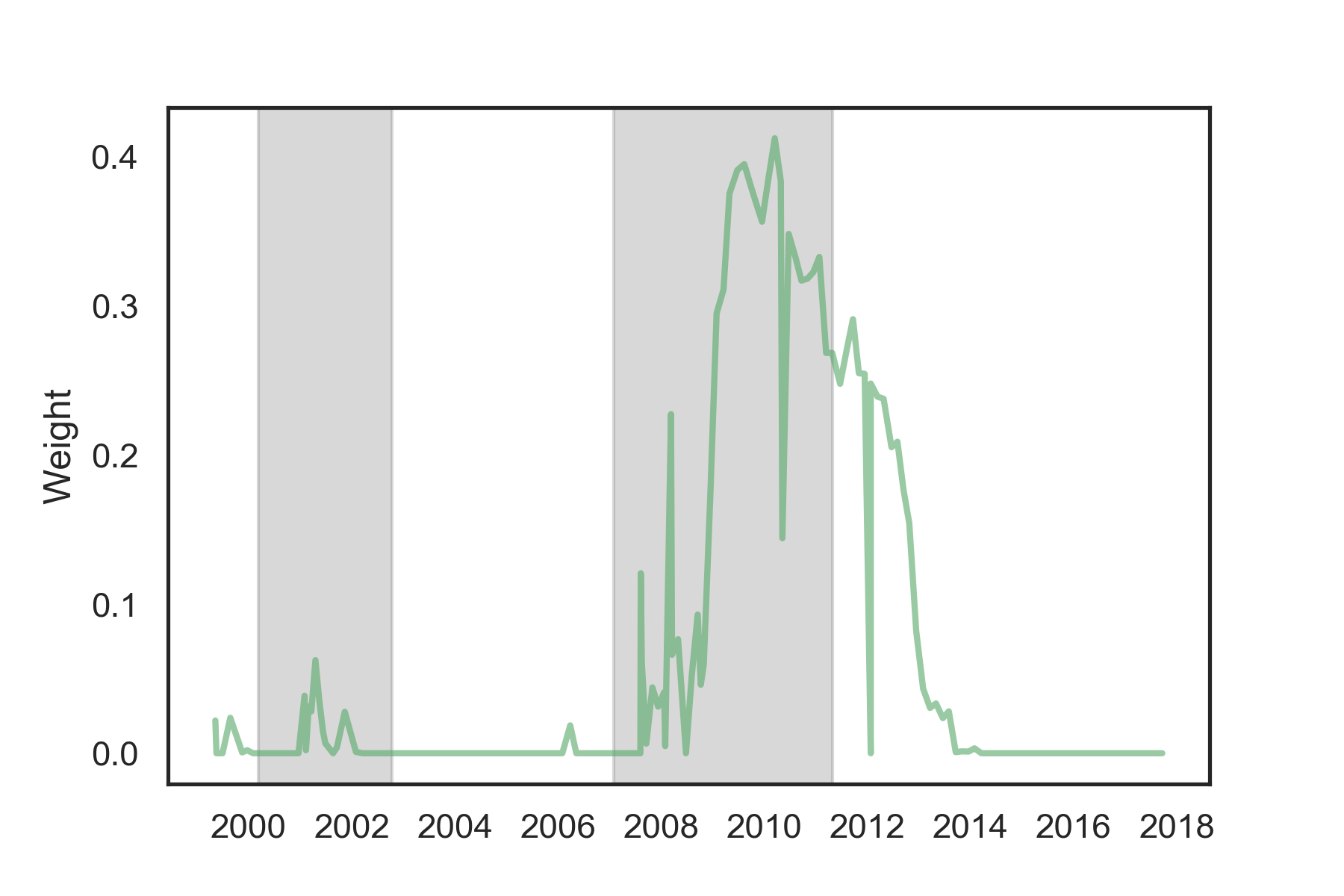}
\label{fig:nmfweights3}
\floatfoot{Notes: The weights of topic three (i.e. `financial' topic) over each statement is plotted as a time series. These weights correspond to the third column of the $W$ matrix derived from the NMF algorithm. These weights are estimated from the NMF model with $k = 3$ and only unigrams considered. The shaded regions represent the crisis periods present in the sample. From left to right, the first shaded region represents the crisis after the bursting of the Dot Com bubble. The second shaded region represents the global financial crisis.}
\end{center}
\end{figure}

Since the documents are released over time, we can present the weights of the topics on the documents over time. This representation shows us the relevance of topics throughout the sample and can provide qualitative information about the content of the FOMC statements. Plots of the weights for each theme are presented in Figures \ref{fig:nmfweights1}-\ref{fig:nmfweights3}.

In the beginning of the sample, we see that information on monetary policy tools and the intermediate targets of monetary policy have the highest weight, however there is also some discussion related to the mandates and economic stability. There was very minor information about financial markets during the dot-com boom around 2001-2002.

Beginning in 2007, we see another peak of information related to monetary policy, but most importantly we see a drastic increase in Theme 3. This is expected given that the topic relates to financial markets and the global financial crisis. This plot also shows how there was continued but decreasing importance of information related to financial markets and the crisis in the statements up until around 2014. This information was likely related to discussing the residual effects of the crisis and recovery actions.

Finally, the post-crisis part of the sample, from 2011 until recent, shows an increase in information related to the mandates and economic conditions.

\subsection{Latent Dirichlet Allocation}
Latent Dirichlet Allocation (Blei et al., 2003) differs from NMF as it is a probabilistic method of topic modeling. Topics in LDA are also interpreted as weighted combinations of words that relate to the main ``theme," however these are derived through probability distributions. This requires a reliance on making distributional assumptions on the probabilistic share of topics in documents. In a similar exposition of the model as presented in Hansen et al. (2017), we have a collection of $D$ documents that contain $M$ unique words. Like NMF, we have to assume $K$ topics, however, these have a vector of probabilities $\beta_{k} \in \Delta^{M-1}$ over the unique terms present in the documents. Probability distribution selection is essential and must allow for the words to appear in multiple topics with varying probabilities.

\begin{figure}[!h]
\begin{center}
\caption{Latent Dirichlet Allocation (LDA) Plate Diagram for Topic Modelling}
\includegraphics[scale=0.35, trim={0cm 5cm 0cm 5.5cm},clip]{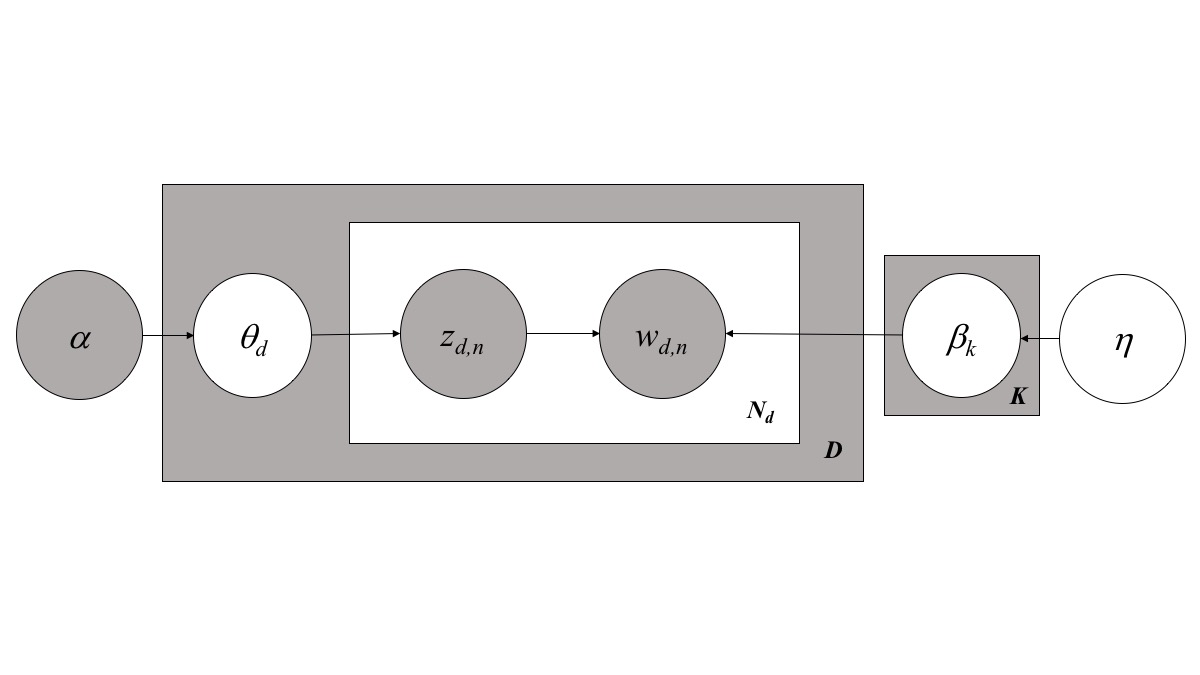}
\floatfoot{Note: The plate diagram shows the flow of parameters within the Latent Dirichelet Allocation model. We show which parameters and their hyperparameters are associated with the document terms ($N_{d}$), the corpus ($D$), and the topics ($K$), and how they are all interconnected to determine topic assignment ($z_{d,n}$) and words present ($w_{d,n}$).} 
\label{fig:LDA}
\end{center}
\end{figure}

LDA allows each document to belong to multiple topics by designating a parameter vector $\theta_{d}$, which represents the distribution over the $K$ topics. This vector is also referred to as the share of the topics in document $d$. Naturally, we will assume that this parameter will follow a Dirichlet distribution. As in Hansen et al. (2017), we assign a symmetric Dirichlet prior with $K$ dimensions and a hyperparameter $\alpha$ to each $\theta_{d}$, as well as a symmetric Dirichlet prior with $M$ dimensions and hyperparameter $\eta$ to each $\beta_{k}$. As noted, the realizations of the Dirichlet distributions with $X$ dimensions lie in the $X-1$ simplex, and the hyperparameters $\alpha$ and $\eta$ control the concentration of the realizations, so that higher values translate into more even probability mass across the dimensions. As in Blei et al (2003), LDA can be viewed as the process in Algorithm \ref{alg:lda}. A plate diagram to understand the structure of the model can be seen in Figure \ref{fig:LDA}.

\begin{center}
\begin{algorithm}[H]
 1. Draw $\beta_{k}$ independently for $k = 1,..., K$ from $Dirichlet(\eta)$\;
 2. Draw $\theta_{d}$ independently for $d = 1,..., D$ from $Dirichlet(\alpha)$\;
 \For{ $w_{d,n}$ in document $d \in D$}{
  (a) Draw a topic assignment $z_{d,n}$ from $\theta_{d}$\;
  (b) Draw $w_{d,n}$ from $\beta_{z_{d,n}}$\;
 }
 \textbf{Note:} scalar values are fixed for hyperparameters $\eta$ and $\alpha$.
 \caption{Latent Dirichlet Allocation}
 \label{alg:lda}
\end{algorithm}
\end{center}

Finally, inference in LDA is tackled by approximating the posterior distributions over $\beta_{k}$ for every $k$ and over $\theta_{d}$ for every $d$ given $K$, $\alpha$, and $\eta$. Following Hansen et al. (2017), we also use the Markov Chain Monte Carlo (MCMC) algorithm by Griffiths and Steyvers (2004). Additionally, we use their default settings for the hyperparameters with $\alpha = 50/K$ and $\eta = 0.025$. As explained by Hansen et al. (2017), the low value of $\eta$ yields a sparse representation of word distributions so topics will have a restricted number of prominent words.

\begin{figure}[!h]
\begin{center}
\caption{NMF vs LDA Coherency}
\includegraphics[scale=0.80]{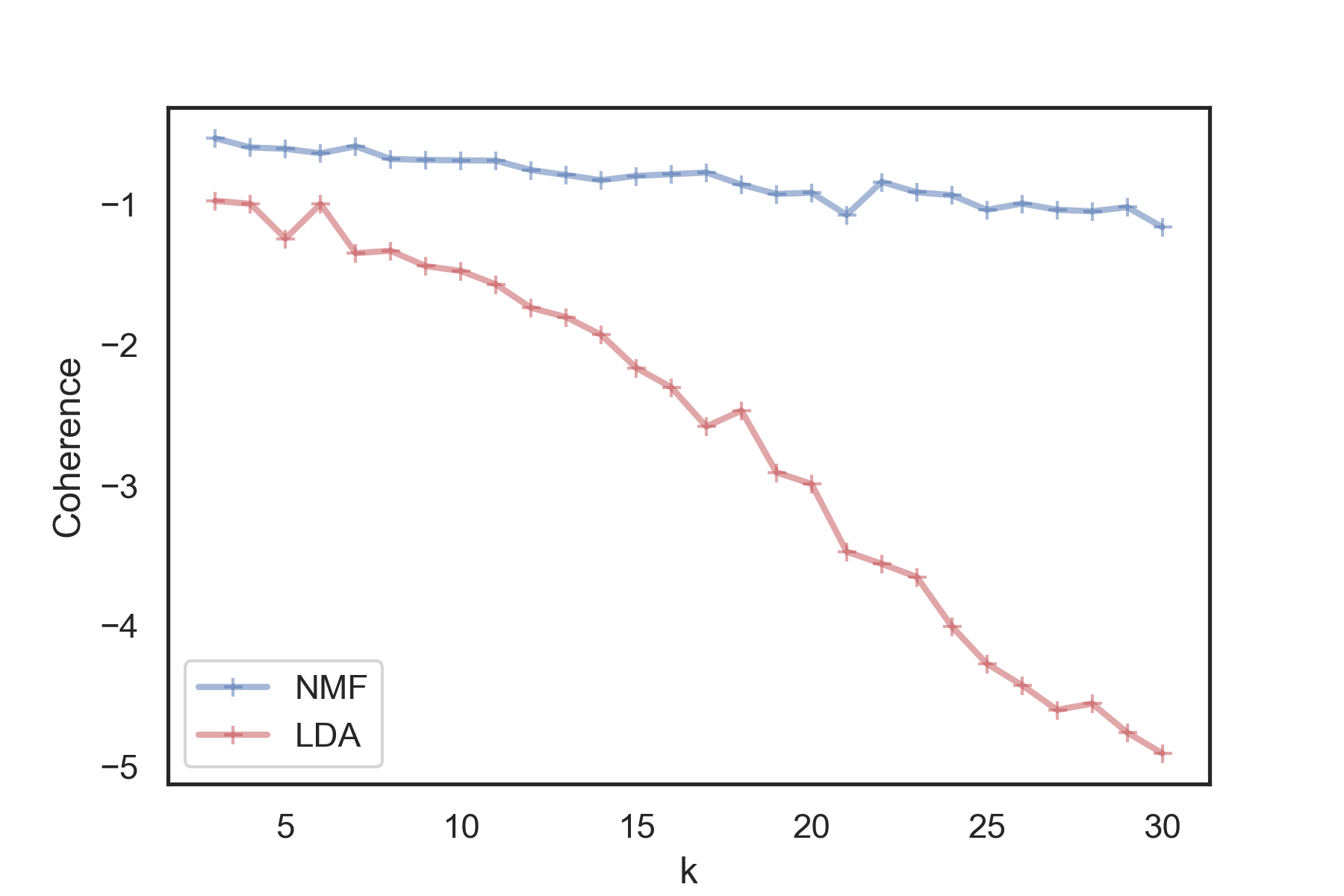}
\label{fig:nmfldacoherence}
\floatfoot{Notes: The coherency of LDA and NMF models are plotted as a function of the number of topics. These coherency values are calculated from a sequence of NMF and LDA models with $k \in [3, 30]$ and only unigrams considered.}
\end{center}
\end{figure}

We present a comparison of the coherency measures between NMF and LDA in Figure \ref{fig:nmfldacoherence}. This graph shows that NMF strictly performs better than LDA across a large range of $k$ values. This is motivation to use NMF for topic modeling of the FOMC statements and proceeding with our analysis using its results.

We note that at lower values of $k$ the difference in coherency between LDA and NMF is smaller. We present the results of estimating the LDA model with $k=3$ in Appendix A. In A.1, we present the top 15 words of each topic. In A.2, we plot the probabilities of documents belonging to that topic as a time series over the documents. Finally, we present the regression results in A.3 for the full sample.

As signaled by the coherency measures, the three-topic LDA model yields less coherent topics. The first topic includes words such as `basis point',`inflation', `security', `growth', and `term'. These words all come from different topics on monetary policy in its traditional sense. `Inflation' and `growth' are considered target of monetary policy, while basis point are associated with the Federal Funds Rate, which is a monetary policy tool itself. In topic 2, we see a mix of terms originating from different aspects of central bank communication, such as `inflation', being a target, and `housing' being a part of financial markets. Finally, theme 3 suffers from the same issues as the first two topics, being composed of words originating from language related to different areas of the monetary policy realm. Despite the incoherence of the topics, we use these topics in the regression analysis as a robustness check.

We would like to note that there is a issue with using LDA model with a low number of topics for regression in this context. This comes from the fact that we sample over a probability simplex with the Dirichlet distribution for the topic share vector $\theta_{d}$. This distribution assumes that all probabilities within this vector will sum to one. This is an assumption that is not made for the weights of the NMF model. Naturally, this brings an issue of the data set of probabilities being a linear function of each other. Therefore, if all topics are included in OLS regression, the $X$ matrix will not be of full column rank and there would be noninvertibility issues due to this presence of multicollinearity. Therefore, we must estimate the regression models based on pairs of themes.

Another issue with regression results in LDA with a low number of topics is the variation of the probabilities over the entire time series. We can see the plots for these in section 2 of Appendix A (Figures \ref{fig:ldaprobabilities1} - \ref{fig:ldaprobabilities3}). There is only a high amount of co-variation for all three topics during the financial crisis of 2007-2011. Outside of this, we normally see large movements in two topics. Since there is only variation in the two topics in this case, we know that changes in the probabilities are one-to-one. This would suggest that we should see symmetric effects within the regression results. These suspicions are confirmed in the regression results reported in section 3 of Appendix A, Tables \ref{table:LDARegression} and \ref{table:LDACrisisRegression}. Therefore, we believe that it is better to make inference using the NMF model.

\subsection{Diebold-Li Model}
A standard representation for the yield curve is the Nelson and Siegel (1987) functional form, which is a three-component exponential approximation. The forward-rate curve in their paper follows the following functional form:
\begin{equation}
\textit{f}_{t}(\tau) = \beta_{1,t} + \beta_{2,t}e^{-\lambda_{t}\tau} + \beta_{3,t}\lambda_{t}e^{-\lambda_{t}\tau}.
\end{equation}
A well-known method, Diebold and Li (2006) extends the above model to 
\begin{equation}
y_{t}(\tau) = \beta_{1,t} + \beta_{2,t}(\dfrac{1-e^{-\lambda_{t}\tau}}{\lambda_{t}\tau}) + \beta_{3,t}(\dfrac{1-e^{-\lambda_{t}\tau}}{\lambda_{t}\tau} - e^{-\lambda_{t}\tau}),
\end{equation}
which is noted to be consistent with the stylized facts about the yield curve. 

The parameter $\lambda_{t}$ represents the exponential decay rate. Small values of $\lambda_{t}$ means that there will be slow decay and can therefore fit the yield curve at longer maturities. Large values of $\lambda_{t}$ produce fast decay and can fit the curve better at short maturities. Additionally, $\lambda_{t}$ governs where the loading on $\beta_{3t}$ achieves its maximum value. In Diebold and Li (2006), $\lambda_{t}$ is set to 0.0609\footnote{As mentioned in Diebold and Li (2006), $\lambda_{t}$ governs the maturity at which the curvature factor is maximized. It is normally assumed that the loading of this factor is maximized between a two- or -three-year maturity. When $\lambda_{t}=0.0609$ the loading of the curvature factor is maximized at 30 months, which is the average between two- and three-year maturities.}. Due to the properties of the exponential terms and the constant as a function of the maturity length ($\tau$), the three latent dynamic factors, $\beta_{1,t}$, $\beta_{2,t}$, $\beta_{3t}$, are interpreted as the ``level", ``slope", and ``curvature" of the yield curve. The loading on $\beta_{1,t}$ is 1, a constant that does not decay to zero in the limit, therefore the factor can be viewed as a long-term factor. The loading on $\beta_{2,t}$ is $(1-e^{-\lambda\tau})/\lambda_{t}\tau$ is a function that starts at 1 but decays monotonically and quickly to 0, hence the factor loads highly on short maturities and is therefore interpreted as the short-term factor. The loading on $\beta_{3,t}$ is $((1-e^{-\lambda_{t}\tau})/\lambda_{t}\tau)-e^{-\lambda_{t}\tau}$ starts at zero, increases, and then decays to zero and therefore the factor is interpreted as the medium-term factor. This is confirmed using empirical proxies of these factors.

Diebold and Li (2006) offer a set of empirical proxies for the level, slope, and curvature factors of the yield curve. For the empirical proxy of the level, it is suggested to use an average of short-, medium-, and long-term yields such as $(y_{t}(3) + y_{t}(24) + y_{t}(120))/3$, although other studies use the maximum maturity yield within their data set as an empirical proxy for the level (such as Hännikäinen (2007)). Their suggested empirical slope is $y_{t}(3) - y_{t}(120)$. Finally, the empirical curvature proxy they use is $2y_{t}(24) - y_{t}(3) - y_{t}(120)$.

One main issue we would like to address with using these empirical proxies for our study is the differing range of maturities of our bond yields. As stated earlier, this larger range is advantageous for summarizing the yield curve, however the empirical proxies may not capture the entire yield curve that we are attempting to explain. Therefore, we modify the equations to be more suitable to our data set. The following empirical proxies are used for our data set:
$$Level_{t} = y_{t}(360)$$
$$Slope_{t} = y_{t}(3) - y_{t}(360)$$
\begin{equation}
Curvature_{t} = 2y_{t}(36) - y_{t}(3) - y_{t}(360)
\end{equation}

\begin{figure}[!h]
\begin{center}
\caption{Level (Long-Term Factor)}
\includegraphics[scale=0.60]{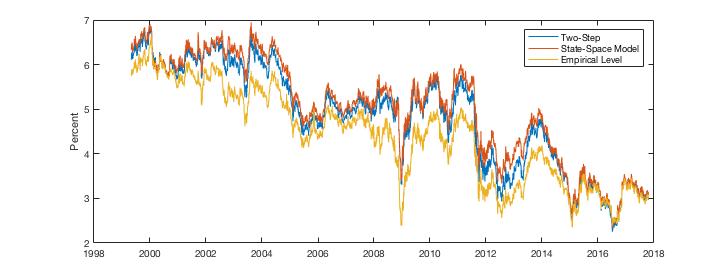}
\label{fig:DLLevel}
\floatfoot{Notes: This plot shows the level factor estimated from the Diebold-Li model with both Two-Step OLS and Kalman filtering and smoothing of the state space model. These are both compared to the empirical proxy of the level of the yield curve.} 
\end{center}
\end{figure}

\begin{figure}[!h]
\begin{center}
\caption{Slope (Short-Term Factor)}
\includegraphics[scale=0.60]{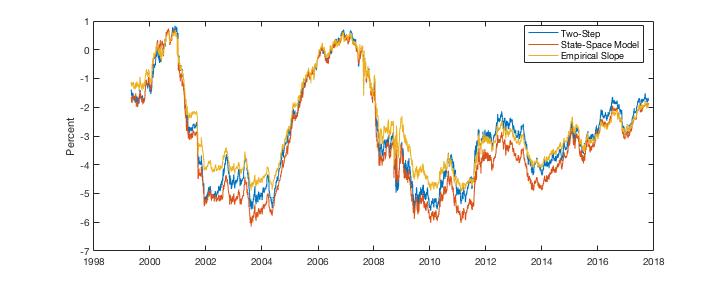}
\label{fig:DLSlope}
\floatfoot{Notes: This plot shows the slope factor estimated from the Diebold-Li model with both Two-Step OLS and Kalman filtering and smoothing of the state space model. These are both compared to the empirical proxy of the slope of the yield curve.} 
\end{center}
\end{figure}

\begin{figure}[!h]
\begin{center}
\caption{Curvature (Mid-Term Factor)}
\includegraphics[scale=0.60]{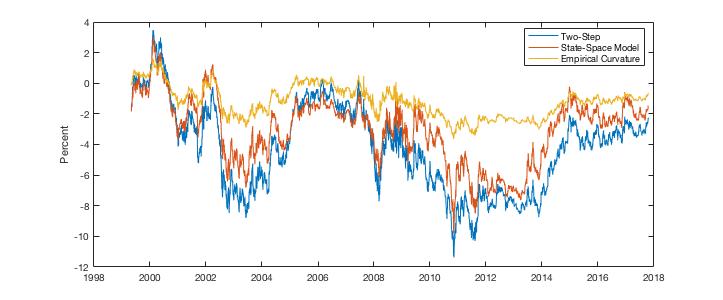}
\label{fig:DLCurvature}
\floatfoot{Notes: This plot shows the curvature factor estimated from the Diebold-Li model with both Two-Step OLS and Kalman filtering and smoothing of the state space model. These are both compared to the empirical proxy of the curvature of the yield curve.} 
\end{center}
\end{figure}

The empirical proxy for the level we use follows the many previous studies that use the maximum maturity yield within the data set. It is a natural yield to use given that the level looks to explain the long term information of the yield curve. The empirical slope we use follows the intuition of taking the difference between the shortest maturity yield and the longest maturity yield. For our data set, this is the difference between the 3-month treasury yield and the 30-year yield. Lastly, our curvature follows the Diebold and Li (2006) empirical proxy closely taking twice a mid-term maturity, for which we use the three-year, and subtracting the shortest maturity yield and the longest maturity yield. We present the comparison of these empirical proxies with our estimated factors in Figures \ref{fig:DLLevel}-\ref{fig:DLCurvature}.

The model is initially estimated in the manner of Diebold and Li (2006) by performing OLS estimation on the above equation for each observation. We use the three estimated factors and estimate a VAR(1) system via OLS. The model is then transformed into a linear state space framework using the output of the OLS estimations as input for initializing the parameters of the Kalman Filter. We then use the smoothed series (after Kalman smoothing) as the measures of the factors.

The state space formulation of the Diebold-Li model is derived from the fact that the factors form a first-order vector autoregressive process. This fact, combined with the original equation, allows us to rewrite the model as a linear state-space system. The state equation is derived directly from the representation of the VAR(1) system:
\begin{equation}
\begin{pmatrix} L_{t}-\mu_{L} \\ S_{t}-\mu_{S} \\ C_{t}-\mu_{C} \end{pmatrix} = \begin{pmatrix} a_{11} & a_{12} & a_{13} \\ a_{21} & a_{22} & a_{23} \\ a_{31} & a_{32} & a_{33} \end{pmatrix} \begin{pmatrix} L_{t-1}-\mu_{L} \\ S_{t-1}-\mu_{S} \\ C_{t-1}-\mu_{C} \end{pmatrix} + \begin{pmatrix} \eta_{t}(L) \\ \eta_{t}(S) \\ \eta_{t}(C) \end{pmatrix}.
\end{equation}

The measurement equation is written as
\begin{equation}
\begin{pmatrix}  y_{t}(\tau_{1}) \\ y_{t}(\tau_{2}) \\ \vdots \\ y_{t}(\tau_{N}) \end{pmatrix} = \begin{pmatrix} 1 & \dfrac{1 - e^{-\lambda\tau_{1}}}{\lambda\tau_{1}} & \dfrac{1 - e^{-\lambda\tau_{1}}}{\lambda\tau_{1}} - e^{-\lambda\tau_{1}} \\ 1 & \dfrac{1 - e^{-\lambda\tau_{2}}}{\lambda\tau_{2}} & \dfrac{1 - e^{-\lambda\tau_{2}}}{\lambda\tau_{2}} - e^{-\lambda\tau_{2}} \\
\vdots & \vdots & \vdots  \\ 1 & \dfrac{1 - e^{-\lambda\tau_{N}}}{\lambda\tau_{N}} & \dfrac{1 - e^{-\lambda\tau_{N}}}{\lambda\tau_{N}} - e^{-\lambda\tau_{N}} \end{pmatrix} \begin{pmatrix} L_{t} \\ S_{t} \\ C_{t} \end{pmatrix} + \begin{pmatrix} e_{t}(\tau_{1}) \\ e_{t}(\tau_{2}) \\ \vdots \\ e_t(\tau_{N}) \end{pmatrix}
\end{equation}

In vector-matrix notation, the state-space system can be represented as  
$$(f_{t} - \mu) = A(f_{t-1}-\mu) + \eta_{t}$$
\begin{equation} 
y_{t} = \Lambda f_{t} + e_{t},
\end{equation}
where $\eta_{t}$ and $e_{t}$ are orthogonal, Gaussian white noise processes defined as 
\begin{equation}
\begin{pmatrix} \eta_{t} \\ e_{t} \end{pmatrix} \sim WN\begin{pmatrix} \begin{pmatrix} 0 \\ 0 \end{pmatrix}, \begin{pmatrix} Q & 0 \\ 0 & H \end{pmatrix} \end{pmatrix}
\end{equation}

As in Diebold and Li (2006), for least-squares optimality of the Kalman filter, it is required to have the orthogonality conditions of the transition and measurement disturbances as well as $\eta_{t}$ and $\epsilon_{t}$ being orthogonal to the initial state vector:
$$E(f_{0}\eta_{t}')=0,$$
\begin{equation}
E(f_{0}\epsilon_{t}')=0.
\end{equation}

\subsection{Regression Analysis}
In order to assess the effect of the themes on movements in the U.S. treasury yield curve, we use standard regression analysis. To identify the impact of the statements on the yield curve, we look at the one-day absolute change in the yield curve factors extracted by the Kalman filter using the Diebold-Li model. We use the absolute change in the factors for two main reasons: first, it provides us with a measure of volatility in the yield curve. This measure of volatility is advantageous in understanding reactions to news since it does not take into account directional information, but simply movements in the different dimensions of the yield curve. This is a method that has also been used by Mazis and Tsekrekos (2017). Our second reason is that our measures of information do not take into account the polarity of information (i.e. ``positivity" or ``negativity"), therefore we can not specify any directional changes. We believe that in order to study the direction of influence, we would have to understand whether not only a statement is negative or positive, but whether information within a topic is negative or positive. Analysis related to polarity falls under the branch of natural language processing called ``sentiment analysis," and unfortunately the entirety of this domain of research depends on dictionary methods for establishing polarity. This is an additional source of subjectivity that we wish to avoid in this analysis.

The baseline estimation equation will take the following form:
\begin{equation}
| \Delta y_{t}| = \beta_{0} + \sum_{k=1}^{3}\beta_{k}\Delta Theme_{k,t} + \delta_{1}TermSpread_{t} + \delta_{2}CreditSpread_{t} + \delta_{3}VIX_{t} + \epsilon_{t},
\end{equation}
with $\Delta y_{t}$ equal to $y_{t}-y_{t-1}$. $y_{t}$ represents the end-of-day measure of the factor on day $t$, while $y_{t-1}$ is equal to the end-of-day measure of the factor on the previous day. Therefore, we are looking at the absolute change of the factor from the previous day.

The $\beta_{k}$ coefficients are the marginal impact of the $K$ themes on the movements in the respective estimated factors related to the yield curve. These are the coefficients of primary interest in our baseline specification. Related to previous research, we study whether these coefficients are significantly different from zero. Related to the studies by Ehrmann and Fratzscher (2007) and Gürkanynak et al. (2005), we hypothesize that there should be a significant impact of the statements on the yield curve with a focus on the medium and long end. However, we cannot say a priori which themes should be of significance. We include the term spread, credit spread, and VIX as controls for reasons described in the data section.

There may be potential issues lying within the baseline specification. First, there is the possibility that statements are particularly important during a crisis and information outside of the crisis is not as influential as previously found. In this case, we may find that there is a significant change in the yield curve on release dates of the statement and this may naturally be attributed to the measured themes. This may spuriously reveal statistical significance of themes. We are explicitly interested in the financial crisis beginning in 2007 as this was a time of extreme market uncertainty and turmoil, which made monetary policy a central focus of economic and financial discussion and therefore made the statements an important source of information to set expectations. To address this, we introduce a crisis dummy variable within the regression as follows:
\begin{equation}
| \Delta y_{t}| = \beta_{0} + \sum_{k=1}^{3}\beta_{k}\Delta Theme_{k,t} + \psi Crisis_{t} + \sum_{k=1}^{3} \eta_{k}\Delta Theme_{k,t} \cdot Crisis_{t} +  \sum_{n=1}^{3}\delta_{n}Control_{n,t} + \epsilon_{t},
\end{equation}
with the $Crisis_{t}$ variable being defined as a dummy variable that is equal to 1 for the days within the crisis period of February 27, 2007 through April 13, 2011 according to the Federal Reserve of St. Louis's Financial Crisis Timeline\footnote{The full timeline can be found at https://www.stlouisfed.org/financial-crisis/full-timeline.} The dummy variable is zero for every other day. The $Control_{n,t}$ variables represent our standard term spread, credit spread, and VIX controls, respectively.

In this specification, the $\beta_{k}$ are still coefficients of interest, however we examine the coefficients on the interaction terms of the change in the themes and the crisis dummy variable ($\eta_{k}$) to see if there is a heterogeneous effect of the impact of the themes on the yield curve factors during the financial crisis. Significant $\eta_{k}$'s would suggest an additional impact of a change in information related to the respective theme during the crisis.

An additional analysis we perform is a sub-sample analysis where we look to see if the results are simply dependent on a particular part of the sample. We split the sample into a pre-financial crisis period (1999-2006) and crisis and post-crisis period (2007-2017). We estimate the baseline regression in both cases, however in the second sub-sample we also allow for a crisis dummy. If we find a significant impact in one sub-sample and not in the other, this may suggest that the effects are time/circumstance dependent. Particularly, if we find a significant impact in the crisis and post-crisis sub-sample and not the pre-crisis sample, it would suggest that unexpected information was more present during the crisis or markets specifically pay more attention to statement releases during the latter sub-sample.

\section{Results}
\subsection{Baseline Results}
In the first part of the analysis, we look at the difference in the mean of the absolute change in the factors on days the FOMC statements are released. In general, we estimate the effects by running a regression of the absolute change in the factor on an event day dummy variable. The results for the full sample are presented in Table \ref{table:event}. The results show that there is a significant difference in the mean on the statement release dates for absolute changes in the level, slope, and curvature. There is a significant, positive increase in the absolute change in the level at the 5 percent level of significance. There is a significant, positive increase in the absolute change in the slope and curvature of the yield curve at the 1 percent level of significance. These results indicate that on the release dates of statements there are significant fluctuations of the yield curve. In general, this supports previous findings that the statements create news that is relevant for financial markets, which is reflected in the changes in the yield curve.

\begin{table}[!h]
\centering
\caption{Statement Release Date Effect, Full Sample}
\begin{tabular}{c c c c}
\hline
\hline
& & & \\
 Dependent Variable & Mean & Event Date Effect & t-stat \\
\hline
\textit{$\Delta$Level} & 0.0133 & 0.0062** & 2.1042\\
& & & \\
\textit{$\Delta$Slope} & 0.0138 & 0.0148*** & 3.6466\\
& & & \\
\textit{$\Delta$Curvature} & 0.0136 & 0.0724*** & 8.2403\\
& & & \\
\textit{N} & 4617 &  &  \\
\hline
\hline
\end{tabular}
\label{table:event}
\begin{tablenotes}
 \small
 \item \textit{Notes:} Significance stars ***, **, * represent significance at the 1, 5, and 10 percent levels, respectively. Coefficients are estimated by running an OLS regression of the absolute changes in the factors on an event date dummy variable. Each regression includes the term spread, credit spread, and VIX as controls.
\end{tablenotes}
\end{table}

We then estimate the baseline regression of the model over the full sample. The results of the OLS regression are presented in Table \ref{table:NMFRegression}. The first regression column of the table represents the regression of the absolute change in the level on the themes and the relevant controls. In this regression, the $\beta_{k}$ are insignificantly different from zero and there is insufficient evidence that there is an impact of any of the themes on changes in the level. We find that the term spread, credit spread, and VIX have a highly significant and positive impact on the fluctuation of the level all at a 1 percent level of significance. The second column of the table represents absolute changes in the slope of the yield curve. Once again, we find that there is no significant impact of the themes on fluctuations in the slope. There is a significant and positive impact of the controls on changes in the slope with the term spread and VIX being significant at the 1 percent level, and the credit spread being significant at the 5 percent level. Finally, the third column of the table looks at the impact of changes in the themes and the controls on the absolute change in the curvature factor of the yield curve. We find that changes in themes 2 and 3 have a significant and positive impact at the 5 percent level of significance, while theme 1 has an insignificant impact. The term spread and VIX are highly significant and have a positive impact at the 1 percent level of significance.

\begin{table}[!h]
\centering
\caption{Baseline OLS results (Full Sample, NMF)}
\begin{tabular}{c c c c}
\hline
\hline
 & (1) & (2) & (3) \\
 Independent Variable & $\Delta$Level & $\Delta$Slope & $\Delta$Curvature \\
\hline
\textit{C} & 0.0135*** & 0.0143*** & 0.0161***\\
&  (0.0015) & (0.0020) & (0.0044)\\
\textit{$\Delta$Theme 1} & 0.1598 & 0.1992 & 0.3073\\
&  (0.1013) & (0.1391) & (0.3035)\\
\textit{$\Delta$Theme 2} & 0.0542 & 0.1328 & 0.5464**\\
& (0.0750) & (0.1029) & (0.2246)\\
\textit{$\Delta$Theme 3} & 0.0386 & 0.1344 & 0.5544**\\
& (0.0794) & (0.1090) & (0.2379) \\
\textit{Term Spread} & 0.0030*** & 0.0013**& 0.0118*** \\
& (0.0004) & (0.0006) & (0.0015)\\
\textit{Credit Spread} & 0.0038*** & 0.0020* & 0.0005\\
& (0.0008) & (0.0011) & (0.0024)\\
\textit{VIX} & 0.0007*** & 0.0016*** & 0.0038***\\
& (0.0001) & (0.0001) & (0.0003)\\
& & & \\
\textit{N} & 4617 & 4617 & 4617 \\
\textit{Adjusted} $R^{2}$ & 0.0944 & 0.1004 & 0.1150 \\
\hline
\hline
\end{tabular}
\label{table:NMFRegression}
\begin{tablenotes}
 \small
 \item \textit{Notes:} Significance stars ***, **, * represent significance at the 1, 5, and 10 percent levels, respectively. Standard errors are reported in parentheses. Each column represents a separate OLS regression estimated for the one-day absolute change in the designated factor extracted from the Diebold-Li model on the changes in the weights from the themes reported in the $W$ matrix of NMF and relevant controls.
\end{tablenotes}
\end{table}

\subsection{Financial Crisis Analysis}
As mentioned before there could be issues with identifying the effect within the whole sample due to exogenous events. In this case, we would like to account for the impact of the financial crisis. We first repeat the above exercise and look to identify whether there is a significant change in the mean of the absolute changes of the factors on days statements are released, however this time we allow for changes in the mean during the crisis as well as changes in the mean for release dates during the crisis. This analysis can be captured by running an OLS regression of the absolute changes in the factors on a crisis year dummy, a release date dummy, and an interaction term of both of the dummy variables. The results are presented in Table \ref{table:eventcrisis}.

In this analysis, we find that there is a highly significant and positive crisis effect of absolute changes in the level, slope, and curvature. This suggests that during the crisis fluctuations of the entire yield curve were higher than outside of the financial crisis. We find that there is no significant change in the mean of the absolute changes in the level and slope on the statement release dates. However, there is a highly significant and positive increase in the mean of the absolute change in the curvature on the statement release dates. Finally, we find that there is evidence of a joint effect in absolute changes in the slope and curvature, statistically significant at the 1 and 5 percent levels, respectively. In the case of the slope, the insignificant overall effect of statement release dates but highly significant joint effect suggests that statements created news relevant to fluctuations in the slope during the crisis. In the case of changes in the curvature, this suggests that there is an additional increase in fluctuations due to statements being released during the crisis.

\begin{table}[!h]
\centering
\caption{Statement Release and Crisis Effects, Full Sample}
\begin{tabular}{c c c c c}
\hline
\hline
& & & &\\
 Dependent Variable & Mean & Event Date Effect & Crisis Effect & Joint Effect\\
\hline
\textit{$\Delta$Level} & 0.0139 & 0.0042 & 0.0025* & 0.0075\\
& & & & \\
\textit{$\Delta$Slope} & 0.0168 & 0.0053 & 0.0144*** & 0.0352***\\
& & & & \\
\textit{$\Delta$Curvature} & 0.0177 & 0.0597*** & 0.0190*** & 0.0471** \\
& & & &\\
\textit{N} & 4617 &  &  &\\
\hline
\hline
\end{tabular}
\label{table:eventcrisis}
\begin{tablenotes}
 \small
 \item \textit{Notes:} Significance stars ***, **, * represent significance at the 1, 5, and 10 percent levels, respectively. Coefficients are estimated by running an OLS regression of the absolute changes in the factors on event date and crisis dummy variables and their interaction. Each regression includes the term spread, credit spread, and VIX as controls.
\end{tablenotes}
\end{table}

We then continue the analysis to determine if there is a significant impact of changes in the themes on fluctuations in the factors and if there is any additional impact during the crisis. We previously determined that there is an overall effect of the release of the FOMC statements on fluctuations in the curvature and a joint effect present in fluctuations in the curvature and slope. Therefore, in this analysis, we look to see if the themes significantly influence any of these impacts. The results are presented in Table \ref{table:NMFCrisisRegression}.

We confirm that the crisis had a significant and positive effect on the absolute changes of the level, slope, and curvature. The impact is significant at the 5 percent level for the level and at the 1 percent level for the slope and curvature. We find that changes in theme 3 have a positive and statistically significant effect on fluctuations in the curvature of the yield curve. This effect is highly significant at the 1 percent level. Themes 1 and 2 have no statistically significant impact on absolute changes in the level, slope, or curvature. This suggests that these themes do not influence changes in the yield curve at all. The regression results also show that the term spread, credit spread, and the VIX are highly significant and have a positive effect at a 1 percent level of significance. We find that the term spread and the VIX have a significant and positive impact on changes in the curvature at a 1 percent level of significance, while only the VIX has a positive and significant impact of absolute changes in the slope at a 1 percent level of significance.

We find that the coefficients on the interaction terms are all insignificantly different from zero. This suggests that there is no additional impact of changes in the themes during the crisis period. Therefore, the significance found of changes in Theme 3 on fluctuations in the curvature of the yield curve is an overall effect and there is no additional impact during the financial crisis.

\begin{table}[!h]
\centering
\caption{Crisis Results (Full Sample, NMF)}
\begin{tabular}{c c c c}
\hline
\hline
 & (1) & (2) & (3) \\
 Independent Variable & $\Delta$Level & $\Delta$Slope & $\Delta$Curvature \\
\hline
\textit{C} & 0.0140*** & 0.0170*** & 0.0196***\\
&  (0.0015) & (0.0020) & (0.0044)\\
\textit{$\Delta$Theme 1} & 0.1173 & 0.1549 & 0.5969\\
&  (0.1234) & (0.1682) & (0.3686)\\
\textit{$\Delta$Theme 2} & 0.0999 & 0.1756 & 0.2619\\
& (0.1039) & (0.1416) & (0.3103)\\
\textit{$\Delta$Theme 3} & 0.0387 & 0.0407 & 1.2385***\\
& (0.1309) & (0.1785) & (0.3910) \\
\textit{Crisis} & 0.0028** & 0.0157*** & 0.0210*** \\
& (0.0014) & (0.0019) & (0.0042) \\
\textit{$\Delta$Theme 1 $x$ Crisis} & 0.1810 & 0.1433 & -0.9269 \\
& (0.2207) & (0.3009) & (0.6592)\\
\textit{$\Delta$Theme 2 $x$ Crisis} & -0.1420 & -0.0932 & 0.3464 \\
& (0.1618) & (0.2206) & (0.4832) \\
\textit{$\Delta$Theme 3 $x$ Crisis} & -0.0633 & 0.0932 & -0.9351 \\
& (0.1765) & (0.2406) & (0.5271) \\
\textit{Term Spread} & 0.0029*** & 0.0009 & 0.0114*** \\
& (0.0004) & (0.0006) & (0.0015)\\
\textit{Credit Spread} & 0.0033*** & -0.0007 & -0.0030\\
& (0.0008) & (0.0012) & (0.0025)\\
\textit{VIX} & 0.0007*** & 0.0017*** & 0.0038***\\
& (0.0001) & (0.0001) & (0.0003) \\
& & & \\
\textit{N} & 4617 & 4617 & 4617 \\
\textit{Adjusted} $R^{2}$ & 0.0947 & 0.1127 & 0.1203 \\
\hline
\hline
\end{tabular}
\label{table:NMFCrisisRegression}
\begin{tablenotes}
 \small
 \item \textit{Notes:} Significance stars ***, **, * represent significance at the 1, 5, and 10 percent levels, respectively. Standard errors are reported in parentheses. Each column represents a separate OLS regression estimated for the one-day absolute change in the designated factor extracted from the Diebold-Li model on the changes in the weights from the themes reported in the $W$ matrix of NMF, a crisis year dummy variable, and the interaction of the weight changes and the dummy variable.
\end{tablenotes}
\end{table}

\subsection{Sub-sample Analysis}

\subsubsection{Before Financial Crisis (1999-2006)}

We conduct the analysis of looking at the impact of the release of policy statements on fluctuations in the multiple dimensions of the yield curve before the crisis. This analysis can reveal whether markets reacted to information generated in the policy statements during this period, as well as possibly shed light onto whether the policy statements revealed any new information for markets to react to. The results of this analysis are presented in Table \ref{table:EventFirst}.

The results show that there is an insignificant change in the mean of the fluctuations of the level, slope and curvature of the yield curve. These results suggest that there was no significant fluctuations on statement release dates in this part of the sample. This is an interesting finding that contrast some of the previous studies who find a significant impact on these dates in the early part of the sample.

\begin{table}[!h]
\centering
\caption{Pre-Financial Crisis Statement Release Analysis, 1999-2006}
\begin{tabular}{c c c c}
\hline
\hline
& & & \\
 Dependent Variable & Mean & Event Date Effect & t-stat \\
\hline
\textit{$\Delta$Level} & 0.0252 & 0.0039 & 1.0194\\
& & & \\
\textit{$\Delta$Slope} & 0.0263 & 0.0034 & 0.6418\\
& & & \\
\textit{$\Delta$Curvature} & 0.0271 & 0.0090 & 0.6946\\
& & & \\
\textit{N} & 1936 &  &  \\
\hline
\hline
\end{tabular}
\label{table:EventFirst}
\begin{tablenotes}
 \small
 \item \textit{Notes:} Significance stars ***, **, * represent significance at the 1, 5, and 10 percent levels, respectively. Coefficients are estimated by running an OLS regression of the absolute changes in the factors on an event date dummy variable. Each regression includes the term spread, credit spread, and VIX as controls.
\end{tablenotes}
\end{table}

Despite there being no significant change in the mean on days statements are released, we move to regression analysis to determine if the themes in the statements are still important in determining the fluctuations of the yield curve. Our results suggest that Theme 3 has a highly significant and positive impact on the slope of the yield curve in this part of the sample at a 1 percent level of significance. The term spread and VIX have a highly significant positive influence on fluctuations in the level and curvature of the yield curve before the crisis, while only the VIX has a significant and positive impact on fluctuations in the slope of the yield curve.

\begin{table}[!h]
\centering
\caption{Pre-Financial Crisis Results (1999-2006)}
\begin{tabular}{c c c c}
\hline
\hline
 & (1) & (2) & (3) \\
 Independent Variable & $\Delta$Level & $\Delta$Slope & $\Delta$Curvature \\
\hline
\textit{C} & 0.0254*** & 0.0265*** & 0.0275***\\
&  (0.0022) & (0.0031) & (0.0074)\\
\textit{$\Delta$Theme 1} & 0.0856 & 0.2864 & -0.2622\\
&  (0.2445) & (0.3399) & (0.8196)\\
\textit{$\Delta$Theme 2} & 0.0842 & 0.1417 & 0.0819\\
& (0.1105) & (0.1536) & (0.3703)\\
\textit{$\Delta$Theme 3} & 0.2622 & 1.3844*** & -0.1058\\
& (0.3475) & (0.4830) & (1.1647) \\
\textit{Term Spread} & 0.0020*** & 0.0008 & 0.0123***\\
& (0.0005) & (0.0007) & (0.0017)\\
\textit{Credit Spread} & -0.0026 & -0.0034 & 0.0043\\
& (0.0020) & (0.0028) & (0.0067)\\
\textit{VIX} & 0.0006*** & 0.0012*** & 0.0030***\\
& (0.0001) & (0.0002) & (0.0005)\\
& & & \\
\textit{N} & 1936 & 1936 & 1936 \\
\textit{Adjusted} $R^{2}$ & 0.0201 & 0.0315 & 0.0864 \\
\hline
\hline
\end{tabular}
\label{table:FirstNMFRegression}
\begin{tablenotes}
 \small
 \item \textit{Notes:} Significance stars ***, **, * represent significance at the 1, 5, and 10 percent levels, respectively. Standard errors are reported in parentheses. Each column represents a separate OLS regression estimated for the one-day absolute change in the designated factor extracted from the Diebold-Li model on the changes in the weights from the themes reported in the $W$ matrix of NMF and relevant controls.
\end{tablenotes}
\end{table}

\subsubsection{Financial Crisis and Post-Financial Crisis Period (2007-2017)}

We repeat the above analysis for the period of 2007 to 2017. This period covers the financial crisis and we try to account for heterogeneous effects in our model specification. As before, we first test for significant differences in the mean of the absolute changes in the `level', `slope', and `curvature' factors. This time we must also account for the fact that there may be differences in the means due to the financial crisis as well as an interaction effect. The results are presented in Table \ref{table:EventSecond}.

We find that there is a highly significant and positive increase in the mean fluctuations during the crisis period for the slope and curvature of the yield curve. However there only seems to be statistically significant decrease in fluctuations of the level during the crisis. There is a significant and positive joint effect for the statement release date during the crisis at the 1 percent level of significance for fluctuations in the slope, but there does not exist overall a significant impact for the release dates on fluctuations in the slope. Our results also suggest that there is a significant, positive impact of the release of the statements on fluctuations in the curvature at the 1 percent level of significance.

\begin{table}[!h]
\centering
\caption{Financial Crisis and Post-Crisis Statement Release Analysis, 2007-2017}
\begin{tabular}{c c c c c}
\hline
\hline
& & & &\\
 Dependent Variable & Mean & Event Date Effect & Crisis Effect & Joint Effect\\
\hline
\textit{$\Delta$Level} & 0.0104 & 0.0043 & -0.0053*** & 0.0066\\
& & & & \\
\textit{$\Delta$Slope} & 0.0161 & 0.0075 & 0.0060** & 0.0320***\\
& & & & \\
\textit{$\Delta$Curvature} & 0.0074 & 0.1209*** & 0.0305*** & -0.0133 \\
& & & &\\
\textit{N} & 2681 &  &  &\\
\hline
\hline
\end{tabular}
\label{table:EventSecond}
\begin{tablenotes}
 \small
 \item \textit{Notes:} Significance stars ***, **, * represent significance at the 1, 5, and 10 percent levels, respectively. Coefficients are estimated by running an OLS regression of the absolute changes in the factors on event date and crisis dummy variables and their interaction. Each regression includes the term spread, credit spread, and VIX as controls.
\end{tablenotes}
\end{table}

In this specification, the $\Delta \text{\textit{Theme 2}}$ term can be neglected as there is only variation of this theme during 2007-2009. Therefore, the relevance of this theme within this sub-sample falls within the crisis period and its effect is captured in the interaction term of the crisis period and changes in theme. Including changes in theme 2 and the interaction term introduces perfect collinearity between those to variables and the data matrix $X$ will no longer be of full rank, which will lead to standard non-invertibility issues in OLS estimation. We must drop one of the variables to avoid this issue.

The results of the regression are presented in Table \ref{table:SecondNMFCrisisRegression}. The crisis dummy variable is highly significant for fluctuations in the level, slope, and curvature, however it has a positive effect for the slope and curvature and a negative effect for the level. We find that changes in Themes 2 does not have a significant impact on fluctuations in the yield curve. Changes in Theme 1 show a significant positive impact on fluctuations in the curvature at the 10 percent level of significance. The coefficient of the interaction term of changes and Theme 1 and the crisis period is also significant at a 10 percent level of confidence, however it has a negative sign. Theme 3 has a highly significant and positive effect on fluctuations in the curvature of the yield curve. However, there is a significant and negative impact of the interaction of changes in Theme 3 with the crisis dummy variable. We find that the term spread is highly significant with a positive impact on fluctuations in the level and curvature of the yield curve. The credit spread is only significant for fluctuations in the slope with a negative sign. Finally, the impact of the VIX is statistically significant at the 1 percent level with a positive effect for fluctuations in the level, slope, and curvature of the treasury yield curve. 

\begin{table}[!h]
\centering
\caption{Crisis and Post-Crisis Results (2007-2017)}
\begin{tabular}{c c c c}
\hline
\hline
 & (1) & (2) & (3) \\
 Independent Variable & $\Delta$Level & $\Delta$Slope & $\Delta$Curvature \\
\hline
\textit{C} & 0.0106*** & 0.0163*** & 0.0112*\\
&  (0.0022) & (0.0030) & (0.0064)\\
\textit{$\Delta$Theme 1} & 0.1183 & 0.0975 & 0.8116*\\
&  (0.1480) & (0.2013) & (0.4240)\\
\textit{$\Delta$Theme 2} & -- & -- & -- \\
&  & & \\
\textit{$\Delta$Theme 3} & 0.0168 & -0.1085 & 1.4282***\\
& (0.1508) & (0.2051) & (0.4319) \\
\textit{Crisis} & -0.0051*** & 0.0071*** & 0.0306*** \\
& (0.0017) & (0.0024) & (0.0050) \\
\textit{$\Delta$Theme 1 $x$ Crisis} & 0.1967 & 0.2303 & -1.1707* \\
& (0.2480) & (0.3373) & (0.7105)\\
\textit{$\Delta$Theme 2 $x$ Crisis} & -0.0714 & 0.0453 & 0.6355 \\
& (0.1349) & (0.1835)  & (0.3864) \\
\textit{$\Delta$Theme 3 $x$ Crisis} & -0.0580 & 0.2226 & -1.1176** \\
& (0.1982) & (0.2696) & (0.5678) \\
\textit{Term Spread} & 0.0039*** & 0.0004 & 0.0126*** \\
& (0.0008) & (0.0012) & (0.0026)\\
\textit{Credit Spread} & -0.0008 & -0.0059*** & 0.0018\\
& (0.0011) & (0.0015) & (0.0032)\\
\textit{VIX} & 0.0016*** & 0.0027*** & 0.0031***\\
& (0.0001) & (0.0002) & (0.0004) \\
& & & \\
\textit{N} & 2681 & 2681 & 2681 \\
\textit{Adjusted} $R^{2}$ & 0.1288 & 0.1455 & 0.1446 \\
\hline
\hline
\end{tabular}
\label{table:SecondNMFCrisisRegression}
\begin{tablenotes}
 \small
 \item \textit{Notes:} Significance stars ***, **, * represent significance at the 1, 5, and 10 percent levels, respectively. Standard errors are reported in parentheses. Each column represents a separate OLS regression estimated for the one-day absolute change in the designated factor extracted from the Diebold-Li model on the changes in the weights from the themes reported in the $W$ matrix of NMF, a crisis year dummy variable, and the interaction of the weight changes and the dummy variable. Changes in Theme 2 estimate as there only exists variation in this theme during the crisis.
\end{tablenotes}
\end{table}

\section{Discussion}
Naturally, this line of research is grounded in efficient market theory; it is assumed that unexpected information (i.e. news shocks) and adjustments in public expectations are reflected in financial markets through fluctuations. In our full sample results looking explicitly at changes in the mean fluctuations on the days the statements are released, we find that there are significant differences along all dimensions of the yield curve, however this is later shown to only be substantial for fluctuations in the curvature of the yield the curve. This supports earlier evidence such as that provided by Kohn and Sack (2004), who find that there is relevant information in the statements that even contributes to the volatility of asset prices for various assets.

We find that there is a general increase relative to normal times in fluctuations of the yield curve during the financial crisis. This was expected as financial markets are characteristically known to be more volatile during times of turmoil. What we find interesting is that there still exists a robust increase in fluctuations in the curvature of the yield curve on statement release dates after controlling for this effect. This finding supports the earlier and later evidence by Musard-Gies (2006) and Mazis and Tsekrekos (2017) who find that the U.S. statements can influence the medium end of the yield curve. However, our lack of significance for a significant effect on fluctuations in the level of the yield curve does not support their findings that the statements can influence the long end of the yield curve. The high level of significance of this effect across all estimated models support results by Leombroni et al. (2017) that communication shocks have the most pronounced effect at intermediate maturities, which generates a hump-shaped response in the yield curve. We are able to attribute this to information related to financial markets and the financial crisis.

Based on the sub-sample analysis, we do not find significant differences in the mean of absolute changes of the yield curve factors on days the statements are released during the pre-crisis period. This could be related to focusing on the yield curve, which may not have had substantial changes in the relatively stable early period. This sample included the dot-com boom, however as mentioned by Rosa (2011, 2013), the statements did not settle on their final form until mid-2003, and therefore could not have been viewed as an entirely informative source of information for expectations until after. However, we still find a significant relationship between the financial theme and curvature.

We find most of the influence of communication on the yield curve present in the crisis and post-crisis period sub-sample. Robustness of the influence of changes in information in financial markets and financial crisis talk was established in increasing curvature fluctuations. However, we find that during the crisis there was a significant decrease in fluctuations of the yield curve. This could signal some sort of direction of the statements during this period, which were likely statements attempting to create stability in the markets, including constant repetition and not revealing any information that may increase volatility.

\section{Conclusion}
\subsection{Summary}
Communication is now a standard tool in the central bank's monetary policy toolkit. Theoretically, communication provides the central bank an opportunity to guide public expectations, and it has been shown empirically that central bank communication can lead to financial market fluctuations. However, there has been little research into which dimensions or topics of information are most important in causing these fluctuations. The previous attempts made to address this issue often employ models that do not offer stable results and are not systematic in the choices of important model parameters. We contribute to this discussion by developing a semi-automatic approach to analyze the influence of central bank communication on the yield curve.

We develop a methodology that summarizes the major themes within a collection of documents, automatically selects the best model, and connects the themes to fluctuations in several dimensions of the yield curve using regression analysis. We empirically show that Federal Open Market Committee (FOMC) communication, through their regular statements can be decomposed into three topics by using Non-negative Matrix Factorization: (i) information related to the mandates, (ii) information related to monetary policy tools, and (iii) information related to financial markets. We find that statements have a significant impact on the shape of the U.S Treasury yield curve and are most influential during financial crisis and the effects are mostly present in the curvature of the yield curve.

\subsection{Limitations}
There exists multiple limitations for this methodological approach that could be addressed to get more precise estimates of the impact of communication on the yield curve. These limitations cover measuring coherency and discovering optimal topics.

First, we would like to address that there exists multiple coherency measures. Generally, they fall under two categories: (i) intrinsic measures and (ii) extrinsic measures. Intrinsic measures are coherency measures that look to assess how coherent a topic is using the original corpus that was used to estimate the topic model. Extrinsic measures measure coherency based on an external or background corpus. Intrinsic measures are often criticized for not accounting for irregular word patterns that may exist in the main corpus and therefore these measures would benefit from using an external corpus as a robustness measure (Röder et al., 2015). However, we believe that the statements may not heavily be influenced by this limitation given that they are typically preprocessed and normalized to fit an existing format. Additionally, the language in terms of terminology and focus in the statements is consistent with language across other sources of communication by the Federal Reserve. Furthermore, it is an issue to find a suitable external corpus covering similar topics and subjects to assess the topics on given formatting issues. For example, the minutes would seem like a suitable source, however it is written in a script-like format, which is not conformable to how the policy statements are formatted.

Another issue with this approach is that the topics given by NMF and LDA are static. This is an obvious disadvantage as language is dynamic and changes over time, and therefore assuming a fixed number of words for a given topic is not reflective of actual language evolution. It would be advantageous to approach this question using dynamic topic model such as Blei and Lafferty (2006), a dynamic extension of LDA, or Greene and Cross (2017), a dynamic extension of NMF. The latter method even allows for new topics to arise and old ones to die. However, these methods require substantially longer documents and many more documents to perform well. Therefore, it would be suitable to perform an analysis like this on European Central Bank communication, including the statement and question and answer session.

The final issue we would like to address would be in extracting topics related to the NMF model. Our study focuses on single terms (i.e. unigrams), which assumes \textit{a priori} independence of terms before topic modeling. This methodology does not account for pairings and groupings of words such as `federal funds rate', `monetary policy', `basis point', or `financial crisis'. This is obviously a disadvantage because many of these terms are naturally used together. Therefore, it would be useful to extend the methodology to cover n-gram models with $n \ in [1,3]$ to account for the $n$-pair terms. Unfortunately, this increases the column dimension of the document-term matrix substantially by adding in all adjacent two-word and three-word pairings of terms within the entire corpus. The main disadvantage of this is the increased computational expense in running the models when performing NMF and LDA.

Despite these several limitations, we believe our results are suggestive of the impact of different themes in the FOMC statements on the treasury yield curve. We can improve on our model to get more precise estimates, but this does not invalidate our current suggestive approach.

\subsection{Policy Implication}
We believe that this methodology is advantageous to further understand the interaction between public policy institutions and market participants. Our results suggest that the role of central bank communication becomes important for determining the yield curve, which may later influence economic outcomes. As mentioned by Svensson (2004), ``monetary policy is to a large extent managing expectations." Policy effectiveness not only depends on controlling short-term interest rates, but significantly depends on a central bank's capability of shaping market expectations. In the increasing age of central bank transparency, we believe that an active and meaningful approach to delivering communication could assist with monetary policy implementation.

\newpage
\section{References}
\singlespacing

\noindent Andersson, M., Dillén, H., \& Sellin, P. (2006). Monetary Policy Signaling and Movements in the Term Structure of Interest Rates. Journal of Monetary Economics, 53(8): 1815-55.
\medskip

\noindent Berger, H., Ehrmann, M., \& Fratzscher, M. (2006). Monetary Policy in the Media. ECB Working Paper 679.
\medskip

\noindent Blei, D. M., \& Lafferty, J. D. (2006). Dynamic topic models. Proceedings of the 23rd International Conference on Machine Learning  - ICML ’06, 113–120.
\medskip

\noindent Blei, D. M., Ng, A. Y., \& Jordan, M. I. (2003). Latent dirichlet allocation. Journal of machine Learning research, 3(Jan), 993-1022.
\medskip

\noindent Blinder, A. S., Ehrmann, M., Fratzscher, M., De Haan, J., \& Jansen, D.-J. (2008). Central Bank Communication and Monetary Policy: A Survey of Theory and Evidence. Journal of Economic Literature, 46(4), 910–945.
\medskip

\noindent Boukus, E., \& Rosenberg, J. V. (2006). The Information Content of FOMC Minutes. SSRN Electronic Journal, (February). https://doi.org/10.2139/ssrn.922312
\medskip

\noindent Brinkhuis, J., \& Tikhomirov, V. (2005). Optimization: Insights and Applications. Princeton University Press.
\medskip

\noindent Connolly, E. \& Kohler, M. (2004). News and Interest Rate Expectations: A Study of Six Central Banks. The Future of Inflation Targeting, ed. Christopher Kent and Simon Guttman. Sydney: Reserve Bank of Australia, 108-34.
\medskip

\noindent Diebold, F. X., \& Li, C. (2006). Forecasting the term structure of government bond yields. Journal of Econometrics, 130(2), 337–364. 
\medskip

\noindent Durbin, J., \& Koopman, S. J. (2001). Time-series Analysis by State Space Model. Oxford Statistical Science Series. Oxford University Press, New York.
\medskip

\noindent Ehrmann, M. \& Fratzscher, M. (2007). Communication by Central Bank Committee Members: Different Strategies, Same Effectiveness? Journal of Money, Credit, and Banking, 39(2-3): 509-41.
\medskip

\noindent Gentzkow, M., Kelly, B. T., \& Taddy, M. (2017). Text as Data. NBER Working Paper Series, 23276, 53.
\medskip

\noindent Greene, D., \& Cross, J. P. (2017). Exploring the Political Agenda of the European Parliament Using a Dynamic Topic Modeling Approach. Political Analysis, 25(1), 77–94.
\medskip

\noindent Gürkaynak, R. S., Sack, B., \& Swanson, E. T. (2005). Do Actions Speak Louder Than Words? The Response of Asset Prices to Monetary Policy Actions and Statements. International Journal of Central Banking, 1(1): 55-93.
\medskip

\noindent Gürkaynak, R. S., Sack, B., \& Wright, J. H. (2007). The U.S. Treasury yield curve: 1961 to the present. Journal of Monetary Economics, 54(8), 2291–2304. 
\medskip

\noindent Hännikäinen, J. (2017). When does the yield curve contain predictive power? Evidence from a data-rich environment. International Journal of Forecasting, 33(4), 1044-1064.
\medskip

\noindent Hansen, S., \& McMahon, M. (2016). Shocking language: Understanding the macroeconomic effects of central bank communication. Journal of International Economics, 99, S114–S133. 
\medskip

\noindent Hansen, S., McMahon, M., \& Prat, A. (2017). Transparency and Deliberation within the FOMC: a Computational Linguistics Approach. The Quarterly Journal of Economics. 
\medskip

\noindent Hendry, S. (2012). Central bank communication or the media's interpretation: What moves markets? (No. 2012-9). Bank of Canada Working Paper.
\medskip

\noindent Jansen, D., \& De Haan, J. (2005). Talking Heads: The Effects of ECB Statements on the Euro-Dollar Exchange Rate. Journal of International Money and Finance, 24(2): 343-361.
\medskip

\noindent Kohn, D. L. \& Sack, B. (2004). Central Bank Talk: Does it Matter and
Why?. Macroeconomics, Monetary Policy, and Financial Stability, Ottawa: Bank of Canada, 175-206.
\medskip

\noindent Lee, D. D., \& Seung, H. S. (1999). Learning the parts of objects by non-negative matrix factorization. Nature, 401(6755), 788–91.
\medskip

\noindent Lee, D. D., \& Seung, H. S. (2001). Algorithms for non-negative matrix factorization. Advances in neural information processing systems (pp. 556-562).
\medskip

\noindent Leombroni, M., Vedolin, A., Venter, G., \& Whelan, P. (2017). Central Bank Communication and the Yield Curve. The 44th European Finance Association Annual Meeting (EFA 2017).
\medskip

\noindent Mazis, P., \& Tsekrekos, A. (2017). Latent semantic analysis of the FOMC statements. Review of Accounting and Finance (Vol. 16).
\medskip

\noindent Mergner, S. (2009). Applications of state space models in finance.
\medskip

\noindent Mimno, D., Wallach, H. M., Talley, E., Leenders, M., \& McCallum, A. (2011). Optimizing semantic coherence in topic models. Proceedings of the conference on empirical methods in natural language processing (pp. 262-272). Association for Computational Linguistics.
\medskip

\noindent Musard-Gies, M. (2006). Do ECB’s Statements Steer Short-Term and Long-Term Interest Rates in the Euro-Zone? The Manchester School, 74 (Supplement): 116–39.
\medskip

\noindent Nelson, C. R., \& Siegel, A. F. (1987). Parsimonious Modeling of Yield Curves. The Journal of Business.
\medskip

\noindent O’Callaghan, D., Greene, D., Carthy, J., \& Cunningham, P. (2015). An analysis of the coherence of descriptors in topic modeling. Expert Systems with Applications, 42(13), 5645–5657.
\medskip

\noindent Reeves, R. \& Sawicki, M. (2007). Do Financial Markets React to Bank of England Communication? European Journal of Political Economy, 23(1): 207-27.
\medskip

\noindent Röder, M., Both, A., \& Hinneburg, A. (2015). Exploring the Space of Topic Coherence Measures. Proceedings of the Eighth ACM International Conference on Web Search and Data Mining - WSDM ’15, 399–408. https://doi.org/10.1145/2684822.2685324
\medskip

\noindent Rosa, C. \& Verga, G. (2007). On the Consistency and Effectiveness of Central Bank Communication: Evidence from the ECB. European Journal of Political Economy, 23(1): 146-75.
\medskip

\noindent Rozkrut, M., Rybiński, K., Sztaba, L., \& Szwaja, R. (2007). Quest for Central Bank Communication. Does It Pay To Be “Talkative”? European Journal of Political Economy, 23(1): 67-87.
\medskip

\noindent Stevens, K., Kegelmeyer, P., Andrzejewski, D., \& Buttler, D. (2012). Exploring topic coherence over many models and many topics. Proceedings of the 2012 Joint Conference on Empirical Methods in Natural Language Processing and Computational Natural Language Learning (pp. 952-961). Association for Computational Linguistics.
\medskip

\noindent Svensson, L. E. (2004). Targeting rules vs. instrument rules for monetary policy: what is wrong with McCallum and Nelson? (No. w10747). National Bureau of Economic Research.

\newpage
\begin{appendices}
\section{Latent Dirichlet Allocation Analysis}
\subsection{Visualizing Estimated Topics}
\begin{figure}[H]
\begin{center}
\caption{Top 15 words of Theme 1, LDA}
\includegraphics[scale=0.60]{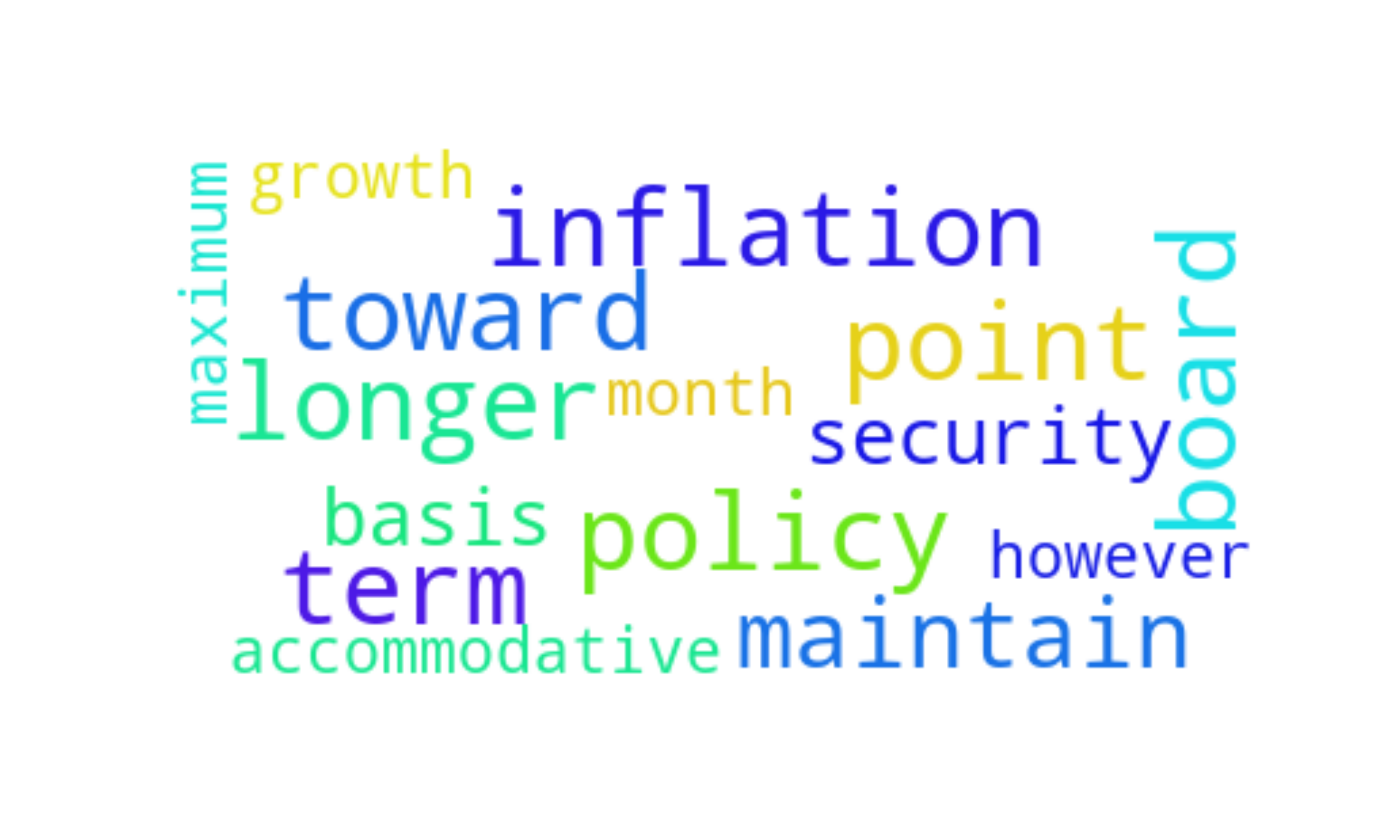}
\label{fig:lda1}
\end{center}
\end{figure}

\begin{figure}[H]
\begin{center}
\caption{Top 15 words of Theme 2, LDA}
\includegraphics[scale=0.60]{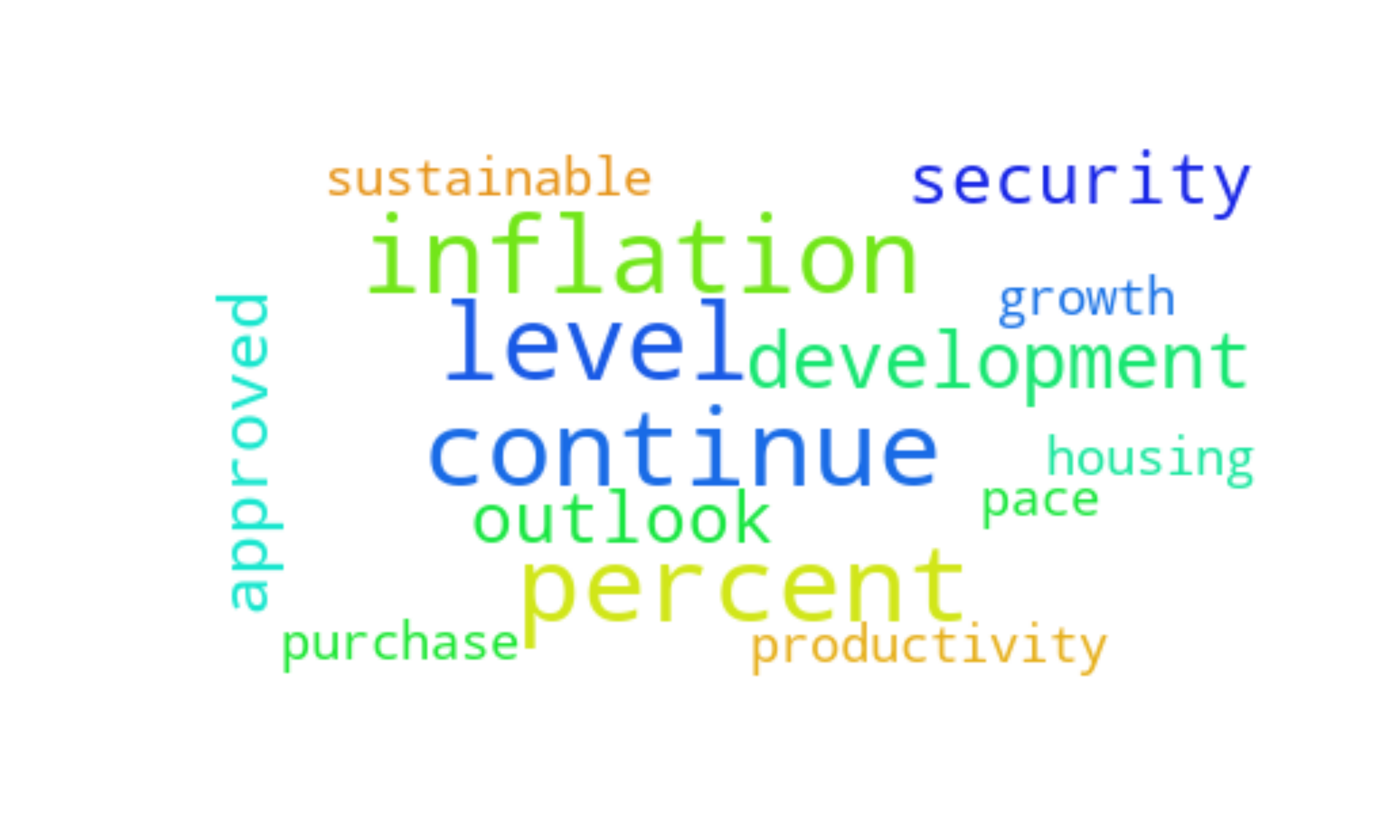}
\label{fig:lda2}
\end{center}
\end{figure}

\begin{figure}[H]
\begin{center}
\caption{Top 15 words of Theme 3, LDA}
\includegraphics[scale=0.60]{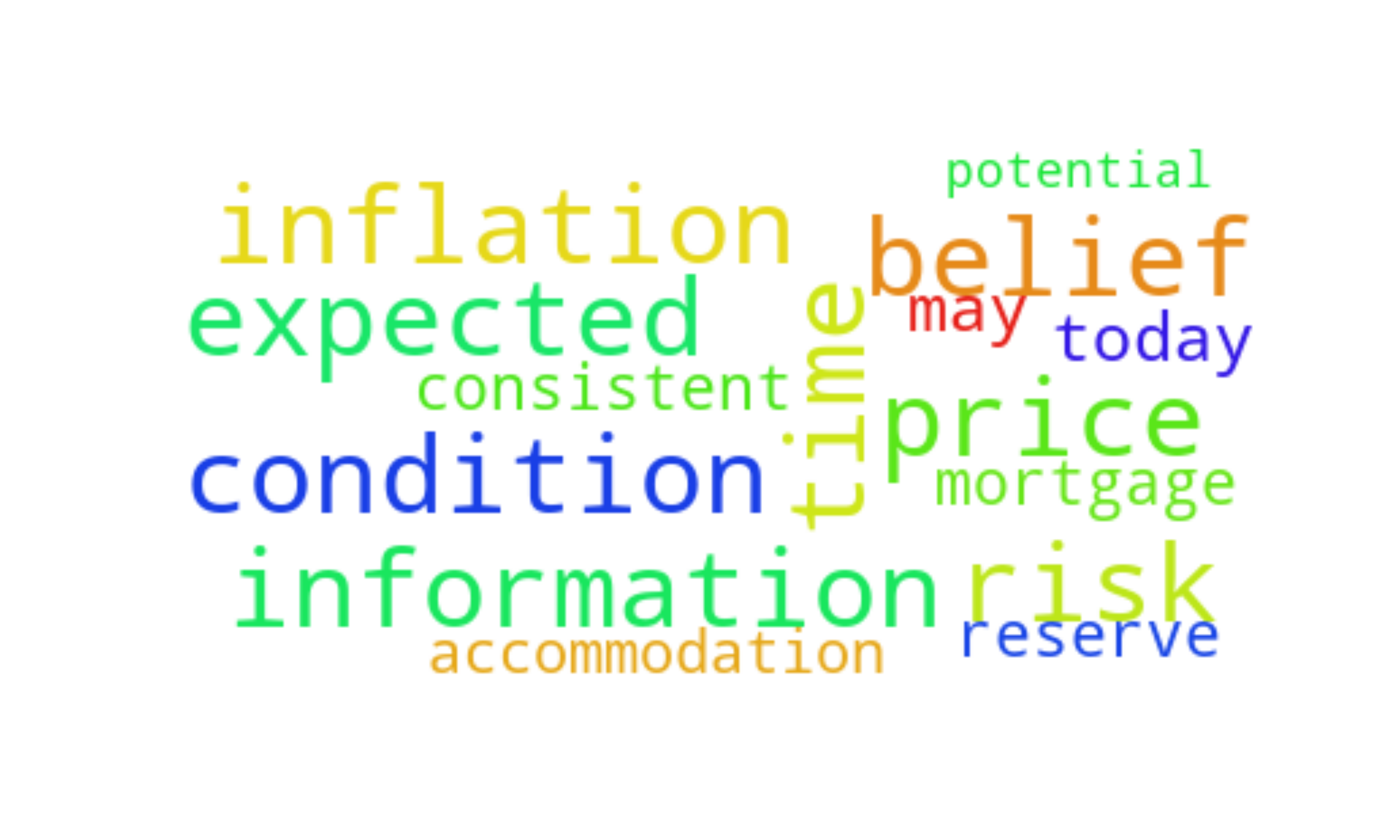}
\label{fig:lda3}
\end{center}
\end{figure}

\subsection{Plots of Probabilities}
\begin{figure}[!h]
\begin{center}
\caption{Theme 1 ('Mandates and Economic Stability') Probabilities, LDA}
\includegraphics[scale=0.80]{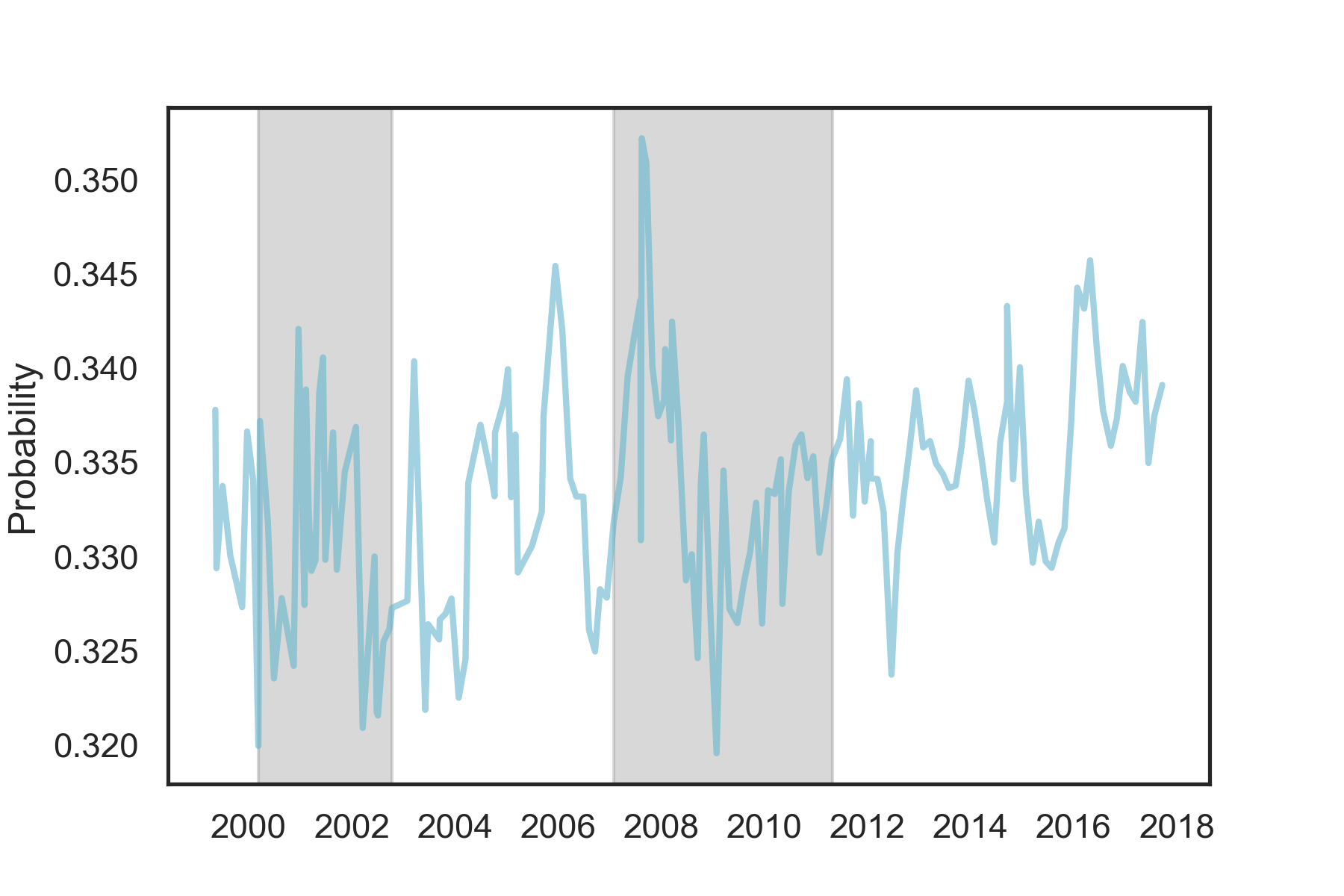}
\label{fig:ldaprobabilities1}
\floatfoot{Notes: The probabilities of topic one over each statement is plotted as a time series. These probabilities correspond to those estimated from the LDA model. These probabilities are estimated from the LDA model with $k = 3$ and only unigrams considered. The shaded regions represent the crisis periods present in the sample. From left to right, the first shaded region represents the crisis after the bursting of the Dot Com bubble. The second shaded region represents the global financial crisis.}
\end{center}
\end{figure}

\begin{figure}[!h]
\begin{center}
\caption{Theme 2 ('Financial Markets \& Financial Crisis') Probabilities, LDA}
\includegraphics[scale=0.80]{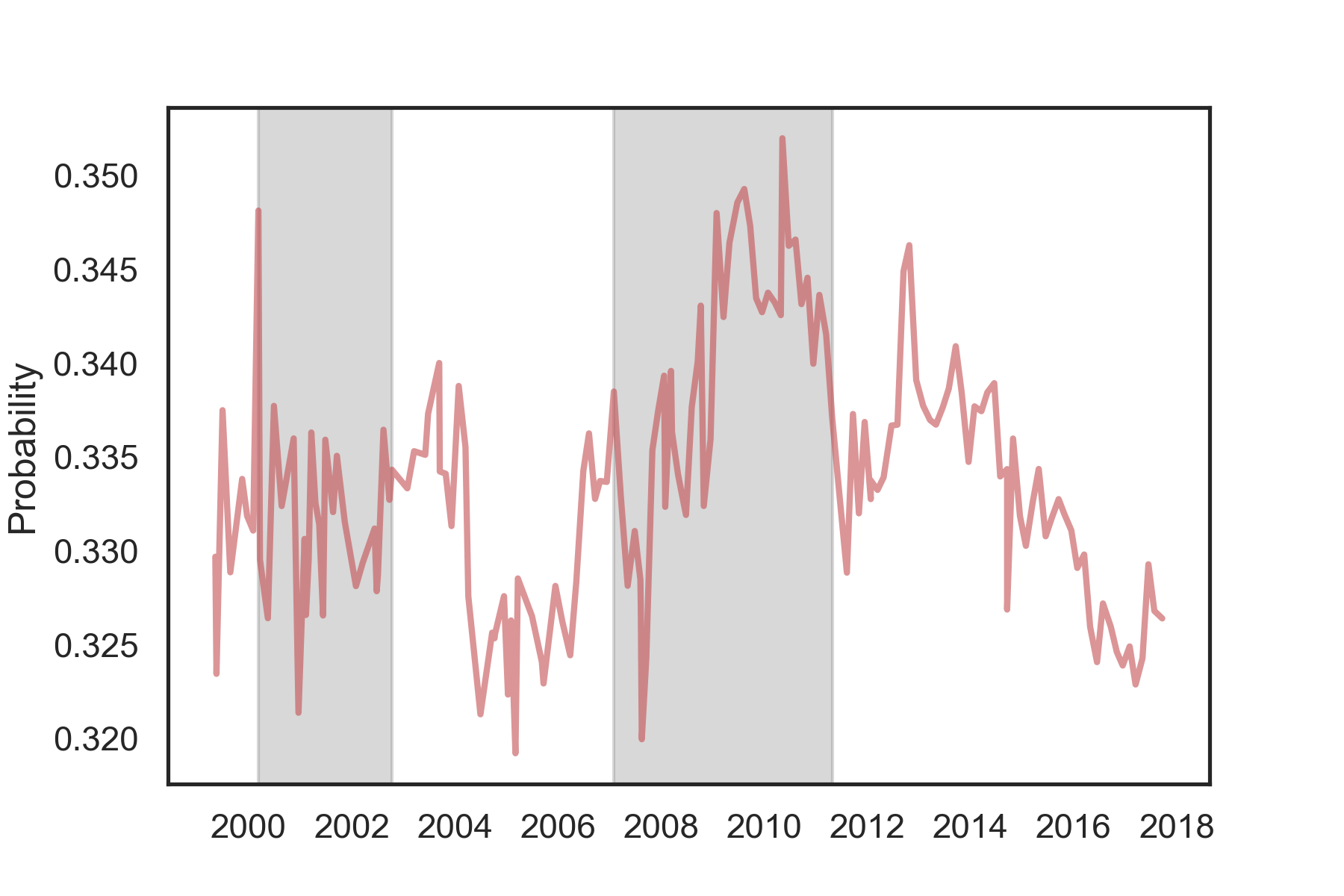}
\label{fig:ldaprobabilities2}
\floatfoot{Notes: The probabilities of topic two over each statement is plotted as a time series. These probabilities correspond to those estimated from the LDA model. These probabilities are estimated from the LDA model with $k = 3$ and only unigrams considered. The shaded regions represent the crisis periods present in the sample. From left to right, the first shaded region represents the crisis after the bursting of the Dot Com bubble. The second shaded region represents the global financial crisis.}
\end{center}
\end{figure}

\begin{figure}[!h]
\begin{center}
\caption{Theme 3 ('Mandates and Economic Stability') Probabilities, LDA}
\includegraphics[scale=0.80]{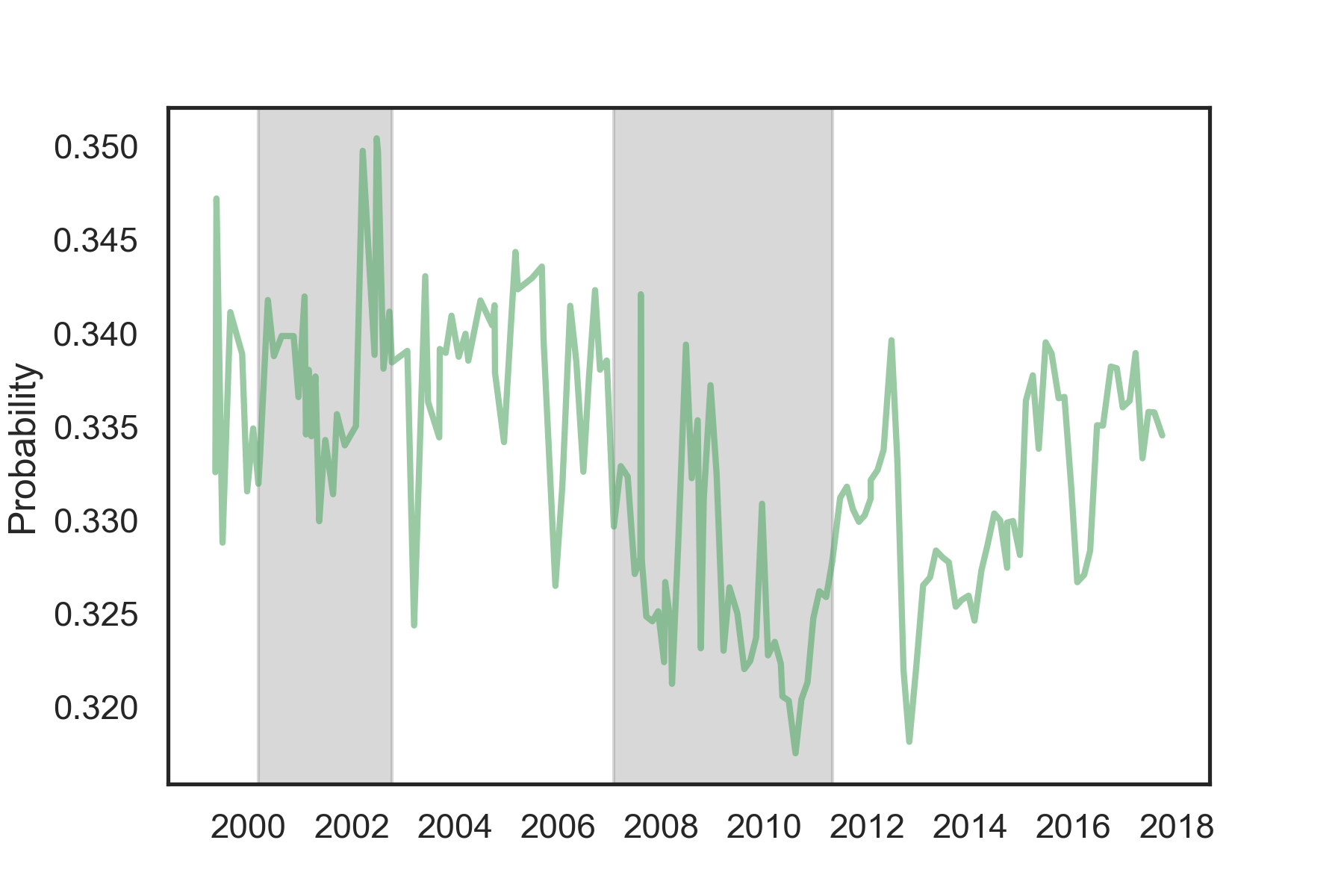}
\label{fig:ldaprobabilities3}
\floatfoot{Notes: The probabilities of topic three over each statement is plotted as a time series. These probabilities correspond to those estimated from the LDA model. These probabilities are estimated from the LDA model with $k = 3$ and only unigrams considered. The shaded regions represent the crisis periods present in the sample. From left to right, the first shaded region represents the crisis after the bursting of the Dot Com bubble. The second shaded region represents the global financial crisis.}
\end{center}
\end{figure}

\begin{landscape}
\subsection{Regression Results}
\begin{table}[!h]
\centering
\caption{Baseline OLS results with LDA, Full Sample}
\begin{tabular}{c c c c c c c c c c}
\hline
\hline
 & (1) & (2) & (3) & (4) & (5) & (6) & (7) & (8) & (9) \\
 Independent Variable & $\Delta$Level & $\Delta$Level & $\Delta$Level & $\Delta$Slope & $\Delta$Slope & $\Delta$Slope & $\Delta$Curvature & $\Delta$Curvature & $\Delta$Curvature \\
\hline
\textit{C} & 0.0135*** & 0.0135*** & 0.0135*** & 0.0142*** & 0.0142*** & 0.0142*** & 0.0158*** & 0.0158*** & 0.0158***\\
&  (0.0015) & (0.0015) & (0.0015) & (0.0020) & (0.0020) & (0.0020) & (0.0044) & (0.0044) & (0.0044)\\
\textit{$\Delta$Theme 1} & 0.0699** & 0.0996** & -- & 0.0383 & 0.0192 & -- & -0.0738 & 0.0603 & --\\
&  (0.0341) & (0.0414) &  & (0.0468) & (0.0568) &  & (0.1022) & (0.1240) & \\
\textit{$\Delta$Theme 2} & -0.0297 & -- & -0.0996** & 0.0191 & -- & -0.0192 & -0.1341 & -- & -0.0603\\
& (0.0334) &  & (0.0414) & (0.0459) &  & (0.0568) & (0.1002) &  & (0.1240)\\
\textit{$\Delta$Theme 3} & -- & 0.0297 & -0.0699** & -- & -0.0191 & -0.0383 & -- & 0.1341 & 0.0738\\
&  & (0.0334) & (0.0341) & & (0.0459) & (0.0468) & & (0.1002) & (0.1022)\\
\textit{Term Spread} & 0.0030*** & 0.0030*** & 0.0030*** & 0.0013** & 0.0013** & 0.0013** & 0.0118*** & 0.0118*** & 0.0118***\\
& (0.0004) & (0.0004) & (0.0004) & (0.0006) & (0.0006) & (0.0006) & (0.0015) & (0.0015) & (0.0015)\\
\textit{Credit Spread} & 0.0038*** & 0.0038*** & 0.0038*** & 0.0020* & 0.0020* & 0.0020* & 0.0007 & 0.0007 & 0.0007\\
& (0.0008) & (0.0008) & (0.0008) & (0.0011) & (0.0011) & (0.0011) & (0.0024) & (0.0024) & (0.0024)\\
\textit{VIX} & 0.0007*** & 0.0007*** & 0.0007*** & 0.0017*** & 0.0017*** & 0.0017*** & 0.0038*** & 0.0038*** & 0.0038***\\
& (0.0001) & (0.0001) & (0.0001) & (0.0001) & (0.0001) & (0.0001) & (0.0003) & (0.0003) & (0.0003)\\
& & & & & & & & & \\
\textit{N} & 4617 & 4617 & 4617 & 4617 & 4617 & 4617 & 4617  & 4617 & 4617\\
\textit{Adjusted} $R^{2}$ & 0.0953 & 0.0953 & 0.0953 & 0.1002 & 0.1002 & 0.1002 & 0.1141 & 0.1141 & 0.1141\\
\hline
\hline
\end{tabular}
\label{table:LDARegression}
\begin{tablenotes}
 \small
 \item \textit{Notes:} Significance stars ***, **, * represent significance at the 1, 5, and 10 percent levels, respectively. Standard errors are reported in parentheses. Each column represents a separate OLS regression estimated for the one-day absolute change in the designated factor extracted from the Diebold-Li model on the changes in the probabilities from the themes found by LDA.
\end{tablenotes}
\end{table}
\end{landscape}

\begin{landscape}
\begin{table}[!h]
\centering
\caption{Crisis Results with LDA, Full Sample}
\begin{tabular}{c c c c c c c c c c}
\hline
\hline
 & (1) & (2) & (3) & (4) & (5) & (6) & (7) & (8) & (9) \\
 Independent Variable & $\Delta$Level & $\Delta$Level & $\Delta$Level & $\Delta$Slope & $\Delta$Slope & $\Delta$Slope & $\Delta$Curvature & $\Delta$Curvature & $\Delta$Curvature \\
\hline
\textit{C} & 0.0140*** & 0.0140*** & 0.0140*** & 0.0170*** & 0.0170*** & 0.0170*** & 0.0194*** & 0.0194*** & 0.0194***\\
&  (0.0015) & (0.0015) & (0.0015) & (0.0020) & (0.0020) & (0.0020) & (0.0045) & (0.0045) & (0.0045)\\
\textit{$\Delta$Theme 1} & 0.0508 & 0.2529 & -- & -0.0707 & 0.2745 & -- & 0.1215 & 2.2102* & --\\
&  (0.0646) & (0.4167) &  & (0.0881) & (0.5683) &  & (0.1931) & (1.2458) & \\
\textit{$\Delta$Theme 2} & -0.2021 & -- & -0.2529 & -0.3452 & -- & -0.2745 & -2.0886 & -- & -2.2102*\\
& (0.4278) &  & (0.4167) & (0.5834) &  & (0.5683) & (1.2788) &  & (1.2458)\\
\textit{$\Delta$Theme 3} & -- & 0.2021 & -0.0508 & -- & 0.3452 & 0.0707 & -- & 2.0886 & -0.1215\\
&  & (0.4278) & (0.0646) & & (0.5834) & (0.0881) & & (1.2788) & (0.1931)\\
\textit{Crisis} & 0.0027* & 0.0027* & 0.0027* & 0.0157*** & 0.0157*** & 0.0157*** & 0.0211*** & 0.0211*** & 0.0211***\\
& (0.0014) & (0.0014) & (0.0014) & (0.0019) & (0.0019) & (0.0019) & (0.0042) & (0.0042) & (0.0042)\\
\textit{$\Delta$Theme 1 x Crisis} & 0.0236 & -0.1518 & -- & 0.1432 & -0.2373 & -- & -0.3101 & -2.2609* & --\\
&  (0.0762) & (0.4191) &  & (0.1040) & (0.5716) &  & (0.2279) & (1.2530) & \\
\textit{$\Delta$Theme 2 x Crisis} & 0.1755 & -- & 0.1518 & 0.3805 & -- & 0.2373 & 1.9509 & -- & 2.2609*\\
& (0.4291) &  & (0.4191) & (0.5852) &  & (0.5716) & (1.2828) &  & (1.2530)\\
\textit{$\Delta$Theme 3 x Crisis} & -- & -0.1755 & -0.0236 & -- & -0.3805 & -0.1432 & -- & -1.9509 & 0.3101\\
&  & (0.4291) & (0.0762) & & (0.5852) & (0.1040) & & (1.2828) & (1.2530)\\
\textit{Term Spread} & 0.0029*** & 0.0029*** & 0.0029*** & 0.0009 & 0.0009 & 0.0009 & 0.0113*** & 0.0113*** & 0.0113***\\
& (0.0004) & (0.0004) & (0.0004) & (0.0006) & (0.0006) & (0.0006) & (0.0015) & (0.0015) & (0.0015)\\
\textit{Credit Spread} & 0.0033*** & 0.0033*** & 0.0033*** & -0.0007 & -0.0007 & -0.0007 & -0.0029 & -0.0029 & -0.0029\\
& (0.0008) & (0.0008) & (0.0008) & (0.0012) & (0.0012) & (0.0012) & (0.0025) & (0.0025) & (0.0025)\\
\textit{VIX} & 0.0007*** & 0.0007*** & 0.0007*** & 0.0017*** & 0.0017*** & 0.0017*** & 0.0038*** & 0.0038*** & 0.0038***\\
& (0.0001) & (0.0001) & (0.0001) & (0.0001) & (0.0001) & (0.0001) & (0.0003) & (0.0003) & (0.0003)\\
& & & & & & & & & \\
\textit{N} & 4617 & 4617 & 4617 & 4617 & 4617 & 4617 & 4617  & 4617 & 4617\\
\textit{Adjusted} $R^{2}$ & 0.0955 & 0.0955 & 0.0955 & 0.1129 & 0.1129 & 0.1129 & 0.1193 & 0.1193 & 0.1193\\
\hline
\hline
\end{tabular}
\label{table:LDACrisisRegression}
\begin{tablenotes}
 \small
 \item \textit{Notes:} Significance stars ***, **, * represent significance at the 1, 5, and 10 percent levels, respectively. Standard errors are reported in parentheses. Each column represents a separate OLS regression estimated for the one-day absolute change in the designated factor extracted from the Diebold-Li model on the changes in the probabilities from the themes found by LDA, a financial crisis dummy, and their interaction.
\end{tablenotes}
\end{table}
\end{landscape}

\section{Optimizing NMF with Local EM Algorithm}
Lee and Seung (2001) provide a multiplicative update rule to solve the Non-negative Matrix Factorization problem. As mentioned before, we look to minimize the objective function of
$$\dfrac{1}{2}||A - WH||_{F}^{2} = \dfrac{1}{2} \sum^{n}_{i=1}\sum^{m}_{j=1}(A_{ij} - (WH)_{ij})^{2}.$$

The algorithm they use to solve this optimization problem is to initialize $W$ and $H$ as non-negative matrices. Then we update the values in $W$ and $H$ by computing the following values until the $W$ and $H$ matrices have converged or the change from the previous matrices fall within a specified tolerance level:

$$H_{[i,j]}^{n+1} \leftarrow H_{[i,j]}^{n}\dfrac{((W^{n})^{T}A)_{[i,j]}}{((W^{n})^{T}W^{n}H^{n})_{[i,j]}}$$

$$W_{[i,j]}^{n+1} \leftarrow W_{[i,j]}^{n} \dfrac{(A(H^{n+1})^{T})_{[i,j]}}{(WH^{n+1}(H^{n+1})^{T})_{[i,j]}}$$

These updates are made on an element by element basis and not matrix multiplication. Each update consists of multiplication by a factor. It intuitive that the $W$ and $H$ multiplicative factor is the identity matrix when $A = WH$. This is the case to that perfect reconstruction is necessarily a fixed point of the update rules (Lee and Seung, 2001). 

Under these update rules, the Euclidean distance $||A - WH||$ is non-increasing. A detailed proof of convergence of this algorithm can be found in Lee and Seung (2001).

\section{Estimating the Diebold-Li Model as a State Space Model}
The Diebold-Li three-factor model can be represented in as a linear state space model. Using this set up, we can use the Kalman filter and Kalman smoothing to get optimal state estimates. The states variables in this model are the `level', `slope', and `curvature' factors that we are interested in. Finally, assuming a Gaussian structure of the errors in the model, we can use standard output of the Kalman filter to retrieve the likelihood function and perform maximum likelihood estimation to estimate the parameters of the model. The exposition of this process in this appendix closely follows Durbin and Koopman (2001). We conform to their exposition and present the Kalman filter and smoothing with the following system
$$y_{t} = Z\alpha_{t} + \epsilon_{t},$$
$$\alpha_{t+1} = T\alpha_{t} + R\eta_{t},$$
with
$$\epsilon_{t} \sim N(0,H),$$
$$\eta_{t} \sim N(0,Q),$$
$$\alpha_{0} \sim N(a_{0}, P_{0}),$$
for $t=1, ...,n$.

\subsection{Kalman Filter}
A filtering approach allows us to produce estimates of the states given information within the data set. Kalman filtering makes updates of knowledge about the state vector as an observation $y_{t}$ becomes available. In this case, the Kalman filter assists us in computing optimal yield predictions and prediction errors based on a noisy signal using information up to time $t$.

The goal of the Kalman filter is retrieve the conditional distribution of the following period state vector $\alpha_{t+1}$ for $t = 1, ..., T$, which are the `level', `slope', and `curvature' factors in the Diebold and Li (2006) model, based on the set of yield observations $Y_{t}$. It achieves this by forward recursion which evaluates one-step ahead estimators. The Kalman filter is initialized using the mean $a_{t+1}$ and the covariance $P_{t+1}$ of the state vector. These can be described as
$$a_{t+1} = E(\alpha_{t+1}|Y_{t})$$
$$P_{t+1} = Var(\alpha_{t+1}|Y_{t}).$$

The mean of the conditional distribution $\alpha_{t+1}$ is an optimal estimator of the state vector at time $t+1$. This comes from the fact that it minimizes the mean squared error matrix, $E((\alpha_{t+1} - a_{t+1})(\alpha_{t+1} - a_{t+1})'|Y_{t})$ for all $\alpha_{t+1}.$ Assuming that $\alpha_{t}$ given $Y_{t-1}$ is normally distributed with mean $a_{t}$ and covariance $P_{t}$ , we can show that $a_{t+1}$ and $P_{t+1}$ can be calculated through a sequence of equations, which are known as the Kalman filtering process.
$$a_{t|t} = a_{t} + P_{t}Z'F_{t}^{-1}v_{t},$$
$$a_{t+1} = Ta_{t} + K_{t}v_{t},$$
$$P_{t|t} = P_{t} - P_{t}Z'F_{t}^{-1}ZP_{t},$$
$$P_{t+1} = TP_{t}L_{t}' + RQR',$$
with
$$v_{t} = y_{t} - E(y_{t}|Y_{t-1}) = y_{t} - E(Z_{t}\alpha_{t} + \epsilon_{t}|Y_{t-1}) = y_{t} - Z\alpha_{t},$$
$$F_{t} = Var(v_{t}|Y_{t-1}) = ZP_{t}Z' + H,$$
$$K_{t} = TP_{t}'F_{t}^{-1},$$
$$L_{t} = T - K_{t}Z,$$
where $t=1,....T$. $v_{t}$ is the one-step ahead forecast error of $y_{t}$  given $Y_{t-1},$ which signals the new information contained in the latest observation. This innovation term is essential for the updating process in estimating $\alpha_{t+1}$. We combine this with the assumptions of $E(v_{t}|Y_{t-1}) = E(v_{t}) = 0$ and $Cov(y_{\tau},v_{t}) = 0$ for $\tau = 1, ..., t-1$. The $K_{t}$ matrix is referred to as the Kalman gain matrix. The $F_{t}$ matrix is assumed to be non-singular. Once $a_{t|t}$ and $P_{t|t}$ are computed, it is sufficient to predict the state vector $\alpha_{t+1}$ and variance matrix at time $t$ with
$$a_{t+1} = Ta_{t|t},$$
$$P_{t+1} = TP_{t|t}T' + RQR'.$$
We initialize the initial state vector mean $\alpha_{0}$ and variance matrix $P_{0}$ with the mean of the two-step OLS estimates of the factors and the variance matrix yielded from the residuals in the $VAR(1)$ system, respectively.

\subsection{Kalman Smoother}
We use the Kalman smoother after Kalman filtering to get the optimal extractions of the factors. Smoothing performs backward recursion that evaluates the mean and variance of specific distributions given the entire set of observations. The initial states are assumed to be known. Through smoothing, we would like to uncover the conditional smoothed state mean $\hat{\alpha_{t}} = E[\alpha_{t}|Y_{t}]$ and the conditional smoothed state variance $V_{t} = Var[\alpha_{t}|Y_{t}]$ given full information. An advantage of smoothing is that the MSE is smaller for the smoothed estimates since they are based on more information than the filtered estimates. Kalman smoothing for the state vector can be represented in the following sequence of equations:
$$r_{t-1} = Z'F_{t}^{-1}v_{t} + L_{t}'r_{t},$$
$$\hat{\alpha_{t}} = a_{t} + Pr_{t-1},$$
$$N_{t-1} = Z'F_{t}^{-1}Z + L_{t}'N_{t}L_{t},$$
$$V_{t} = P_{t} - P_{t}N_{t-1}P_{t},$$
where $t = T, ..., 1$. The vector $r_{t-1}$ is a weighted sum of the future innovations, and $N_{t}$ is the variance matrix of $r_{t}$. The recursive process is initialized with $r_{T}=0$ and $N_{T}=0$. We make the assumption that $\alpha_{1}$ is normally distributed with mean $a_{1}$ and variance-covariance matrix $P_{1}$. In addition to this for estimating the parameters, we compute smoothed disturbances $\hat{\epsilon_{t}} \sim E(\epsilon | Y_{t})$ and $\hat{v_{t}} \sim E(v_{t} | Y_{t})$. The backward recursion equations for disturbance smoothing are as follows
$$u_{t} = F_{t}^{-1}v_{t} - K_{t}'r_{t},$$
$$D_{t} = F_{t}^{-1} + K_{t}'N_{t}K_{t},$$
$$\hat{\epsilon_{t}}= Hu_{t},$$
$$Var(\epsilon | Y_{n}) = H - HD_{t}H,$$
$$\hat{v_{t}} = QR'u_{t},$$
$$Var(v_{t} | Y_{n}) = Q - QR'D_{t}RQ,$$
$$N_{t-1} = Z'F_{t}^{-1}Z + L_{t}'N_{t}L_{t},$$
where $t = 1,..., T$.

\subsection{Maximum Likelihood Estimation}
We need to perform maximum likelihood estimation to determine the parameters of the model. To perform maximum likelihood estimation, we need a parametric model that can form a likelihood function in the form of a joint probability density function:
$$L(y,\psi) = p(y_{1}, ..., y_{T}) = \prod_{t=1}^{T}p(y_{t}).$$

A Gaussian likelihood function can be evaluated of the Diebold-Li model by using a prediction error decomposition of the likelihood. The log-likelihood function is as follows
$$logL(\psi) = logp(y_{1},...,y_{T}; \psi) = -\dfrac{Tp}{2}log(2\pi) - \dfrac{1}{2}\sum_{t=1}^{T}(log|F_{t}| + v_{t}'F_{t}^{-1}v_{t}),$$
where $v_{t}$ is defined as above and interpreted as a vector of prediction errors. $F_{t}$ and $v_{t}$ are standard output of the Kalman filter for given values of $\psi$, and therefore we can get the log-likelihood from running the Kalman filter. Initial values come from the two-step OLS estimation of the model. The initial transition matrix $T$ comes from the $VAR(1)$ coefficient matrix in the system of factors and $\lambda$ is set to the suggested value of 0.0609. Finally, all variances are set to 1.

Maximization of this likelihood function can be conducted through suitable optimization algorithms. We use the Broyden-Fletcher-GoldFarb-Shanno (BFGS) method which uses numerical derivatives as a modified Newton's method. Details on optimization using this algorithm can be found in Brinkhuis and Tikhomirov (2005). We use the state space model package in Matlab to estimate the state space model, which uses these methods in a highly efficient and optimized implementation.

\end{appendices}
\end{document}